\documentstyle[aps,pra,epsf,preprint]{revtex} 
\begin{document}
\draft 
%\rm\Large

\title{\LARGE\bf Dynamics of multiply charged ions in intense laser
fields}

\author{\large\rm S.  X.  Hu $^*$ \ and \ C.  H.  Keitel $^{**}$\\}

\address{\large\sl Theoretische Quantendynamik, 
Fakult$\stackrel{..}{a}$t f$\stackrel{..}{u}$r Physik,
Albert-Ludwigs-Universit$\stackrel{..}{a}$t Freiburg, 
Hermann-Herder-Str.3,
D-79104 Freiburg, Germany}

\date{\Large\today}

\maketitle

\begin{abstract} 

We numerically investigate the dynamics of multiply charged hydrogenic ions 
in near-optical linearly polarized laser fields with intensities of order 
$10^{16}$ to  $10^{17} W/cm^2$.  
The weakly relativistic interaction is appropriately described by the Hamiltonian
arising from the expansion of the Dirac equation up to the second order in the ratio of 
the electron velocity $v$ and the speed of light $c$.
% and by neglecting the extremely weak spin-induced forces in the
% dimension of the magnetic field component of the linearly polized laser field.
Depending on the charge state $Z$ of the ion the relation of strength between laser 
field and ionic core changes. We find around $Z=12$ typical multiphoton
dynamics and for $Z=3$ tunneling behaviour, however with clear relativistic signatures. 
In first order in $v/c$ the magnetic field component of the laser field induces a 
Z-dependent drift in the laser propagation direction and a substantial Z-dependent angular momentum 
with repect to the ionic core. While spin oscillations occur already in first order in $v/c$ 
as described by the Pauli equation, spin induced forces via spin orbit coupling only appear in
the parameter regime where $(v/c)^2$ corrections are significant. In this regime for $Z=12$ ions 
we show strong splittings of resonant spectral lines due to spin-orbit coupling and substantial 
corrections to the conventional Stark shift due to the relativistic mass shift while those to 
the Darwin term are shown to be small.
For smaller charges or higher laser intensities, parts of the electronic wavepacket may tunnel
through the potential barrier of the ionic core, and when recombining are shown to give rise to 
keV harmonics in the radiation spectrum. Some parts of the wavepacket do not recombine after 
ionisation and we find very energetic electrons in the weakly relativistic regime of above 
threshold ionization.

\vspace{1cm}


%\vspace{1cm} \noindent \underline{\hspace{6cm}}

\end{abstract}

\newpage

\section {INTRODUCTION}

\indent 

Various techniques to generate ultrashort pulses such as
chirped-pulse-amplification (CPA) \cite{CPA} have been developed and perfected 
over the last years such that nowadays powerful laser pulses are available over a wide 
range of frequencies and pulse lengths up to intensities of  $10^{21}W/cm^2$ \cite{T31}.
For intensities of $10^{16}W/cm^2$ to  $10^{19}W/cm^2$ the electric field strength of the laser
pulse is already comparable up to far stronger than the atomic unit field strength 
($\sim 5.14\times 10^9 V/cm$) that is experienced by an electron on the first Bohr orbit of 
hydrogen and in fact those fields are already accessible in quite numerous laboratories worldwide 
in rather small table-top set-ups.  

Especially for laser intensities till at about  $10^{16}W/cm^2$ there has been a lot of activity over the
last two decades in intense nonrelativistic laser interactions with matter including atoms 
\cite{atoms}, molecules \cite{molecule}, clusters \cite{cluster} and solids \cite{solid}.  
Many highly nonlinear optical phenomena such as high-order harmonic generation 
\cite{harmonic1,hhgrule,harmonic2,harmonic3}, above-threshold ionization (ATI) \cite{ati} 
and for higher frequencies and intensities stabilization \cite{stab}, have attracted much 
attention for both experimentalists and theorists.  
Since optical laser fields of intensity $10^{21}W/cm^2$ have become available in a rather short time,
both theoretical and experimental activities moved quite quickly into the regime of fully
relativistic dynamics up to around $10^{18}W/cm^2$ to  $10^{22}W/cm^2$ \cite{stabrel,rel,dirac} with experimental
emphasis on free electron dynamics, QED effects and nuclear reaction processes \cite{nucl1,freeex}.
There has been rather little activity however for neutral atoms interacting with optical laser fields
of intensity $10^{16}W/cm^2$ to  $10^{17}W/cm^2$ merely for the reason, that there is almost instantaneous
ionization for those intensities and the free electron dynamics is quite well understood for a long time \cite{free}.
For higher frequencies ionization may be retained at those intensities however will also occur eventually for larger
laser intensities due to the magnetic drift \cite{stabrel} with a likelyhood for magnetic recollisions being usually
smaller than that in the laser polarisation direction \cite{David}.  

For an ionic system one may have the unique possibility to apply relativistic near-optical laser 
field pulses and still allow for bound dynamics. The physical processes occuring will not merely 
scale with the ionic charge but be fundamentally different because of relativity and later QED governing 
the dynamics.  Nowadays ions may be processed from essentially all existing atoms with abitrary charge state, 
absolute purity and quite high density by sending them through foils \cite{gsi} while due to lasers high charge
states have also been achieved as well however with limits in the width of the charge distribution and the absolute 
charge \cite{mions}.
The mean electric field sensed by a ground state electron of hydrogenic uranium corresponds to  
$1.8\times 10^{16} V/cm$. Thus laser fields that are comparable in strength need have an intensity 
of order $10^{29}W/cm^2$ and are thus far beyond reach nowadays. For those intensities the dynamics will 
be absolutely remote from that of hydrogen with say a laser field of  $10^{14}W/cm^2$ and we do not aim 
to make the corresponding comparison in this paper. 
We consider it however more interesting to carry out a comparision of the nonrelativistic laser atom 
interaction with the weakly relativistic dynamics of an ion of charge around 10 with laser fields of  
intensity $10^{16}W/cm^2$ to  $10^{17}W/cm^2$ (see figure 1). Here relativity induces merely corrections 
and it is possible to identify the leading deviations to nonrelativistic dynamics as that of the laser 
magnetic field component, the break-down of the dipole approximation, spin-orbit coupling, zitterbewegung 
amd the relativistic mass shift. In earlier letters we have shown that the magnetic field component of the laser
pulse induces an enhanced angular momentum and thus a reduced electron expectation value in the vicinity of the
multiply charged ion \cite{hole} and the existence of spin signatures in bound electron dynamics and 
radiation \cite{spin}.

In terms of applications multiply charged ions in relativistic laser fields appear also attractive for the 
generation of energic electrons and high harmonics. The kinetic energy of a laser-accelerated electron as 
characterised by the parameter $U_p$ does certainly increase with rising laser intensity. For harmonic generation 
especially it is necessary that the highly energetic electron interacts periodically with the nucleus, urging us to 
enhance the charge of the ionic core, i.e.\ the ionization potential $I_p$ correspondingly. As long as tunneling 
and recollisions can be assured with an appropriate relation of $U_p$ and $I_p$ the simultaneous increase of both 
parameters is obviously attractive as the cut-off energy of high harmonic generation is given by the sum of $I_p$ 
and the maximal kinetic energy at the time of recollision being of order $U_p$.

\indent 
In this paper, we study the dynamics of multiply charged ions in intense lasers fields in the weakly relativistic
parameter regime where terms up to $1/c^2$ are still of significance and higher terms become negligible.
For this purpose we solve numerically the two-dimensional time-dependent expansion of the Dirac equation  
which is exact up to all orders in $1/c^2$ and which was first derived by Foldy and Wouthuysen \cite{foldy}. 
Different weakly relativistic effects arise and we can investigate each by solving the dynamics and comparing the 
situations in which we include or neglect the corresponding part of the Hamiltoninan associated with the effect of interest. 
As a function of the ionic charge relative to the laser field intensity the electric and  magnetic field components
of the laser field will be shown to be either strong enough to lead to ionization in the polarization and propagation 
directions or give rise to interesting structures in the near vicinity of the ionic core. 
We point out the role of the spin in inducing stronger binding of the electronic wavepacket and multiphoton spectral 
line splitting due to spin orbit coupling. This is compared with the situation via the Pauli equation where spin induced 
forces are neglected.
We further show that the relativistic mass shift induces a significant shift of especially the highly 
excited eigenenergies of the ion with respect to the conventional Stark shift and point out the consequences 
for the spectral lines.
For relatively high intensities special emphasis is placed on the tunneling regime in the weakly relativistic
regime. For multiply charged ions we find that both tunneling and recollisions are still possible and indicate 
harmonic generation in the keV regime. We note however that the plateau of the harmonic spectrum, as well known 
in the nonrelativistic regime, is tilted increasingly in the relativistic regime with rising charge. 
This deviation to conventional high harmonic spectra is not too surprising as an electric wavepacket attempting to recollide
due to a phase shift of the electric field in the laser polarization direction may still partly or totally miss the ionic 
core due the magnetic drift in the laser propagation direction, especially for long recollision times.  
In the above threshold spectra we find high electron energies in the keV regime which are separated by the 
photon energy of the applied laser field.

\indent This paper is arranged as follows:  In section II, we derive the Hamiltonian of interest along the lines of  
Foldy and Wouthuysen, then describe the numerical methods for solving the corresponding dynamical equation and present the  
details for computing the observable quantities of interest. In subsection III(A) we then present effects up to first order in $1/c$ 
as those induced by the magnetic field component of the laser field. 
In subsection III(B) we investigate the relativistic corrections to the Stark shift for laser-driven ions followed 
by a study of spin induced forces and the consequent splitting of spectral features in III(C). 
Subsequently, relativistic high order harmonic generation is discussed in III(D), as well as the photoelectron spectra in
subsection III(E).  Finally, a conclusion is drawn.

\section{Preliminary Considerations}

In this more technical section we present subsequently the Hamiltonian describing our system of interest, 
explain how we solve the corresponding dynamical equation numerically via the split-step mechanism and 
finally quantify the observables to be investigated in the following section.

\subsection{System and Hamiltonian}

\indent 
We are interested throughout this article in the dynamics of multiply charged ions of charge state up to $Z=12$ 
with moderate laser intensities ($10^{16} \sim 10^{17}W/cm^2$) and near optical frequencies for the KrF 
(248nm; 0.1838 a.u.) laser system and the doubled Nd:glass laser frequency (527nm; 0.0866 a.u.).
The maximum velocity of electron wavepackets reaches the order $v= 0.1c$, where $c=137.036$ denotes the speed of 
light in atomic units. 
We are therefore entitled to consider the Dirac equation up to second order in $v/c$ and have
confirmed that for our parameters high order corrections do not play a role. One may derive this Hamiltonian 
via various unitary transformations along the lines of the Foldy-Wouthuysen (FW) expansion \cite{foldy} of the 
Dirac equation or alternatively, with the same result, find the first order relativistic corrections to
the Pauli equation \cite{Strange}. As opposed to nonrelativistic treatments we need to include at least
two dimensions in the calculation as the magnetic field component of the laser pulse may induce a significant 
drift in the laser propagation direction. There is a spin-induced force in the magnetic field direction 
\cite{spinmarkus}, i.e. in the remaining third dimension, but the influence is small for the observables and the 
parameters of interest here.  

Our working Hamiltonian ${\bf H_{FW}}$ involves a series of relativistic corrections to 
the usual nonrelativistic Schr\"odinger Hamiltonian.  
One well-known correction term in $v/c$ to the Schr\"odinger equation is the additional 
term in the Pauli equation representing the coupling of the laser magnetic field to the
spin degree of freedom of the electron wavepacket. 
The second order terms include the spin orbit coupling, i.e. the feed-back of the oscillating spin
to the electron motion,  as well as the leading relativistic mass shift term and
Zitterbewegung. Terms of order $O(1/c^3)$ do not play a role for the parameters employed though
would certainly be necessary to include for the fully relativistic regime involving more intense
laser fields and higher charged ions. The main advantage
of using this equation in comparison to the full Dirac equation is
the possible isolation of the influence of each physical mechanism arising. In 
addition this equation does not limit us numerically to use high frequency lasers
as necessary so far with the full Dirac equation \cite{dirac}.

For the circumstances with laser parameters described above, the 
FW Hamiltonian ($2\times 2$ matrix) of a 
bound electron in a strong laser field can be written (in
atomic units \cite{au}) as

\begin{equation}
 	\begin{array}{lcl}
	{\bf H_{FW} } & = & {\bf H_0} + {\bf H_p} + {\bf H_{kin}}+
 		{\bf H_{D}}  + {\bf H_{so}}    \\
	{\bf H_0} & = & ({\bf p+A}(z,t)/c)^2 /2 +V(x,z)\\
	{\bf H_P} & = & {\bf \sigma} \cdot {\bf B}(z,t) /2c\\
	{\bf H_{kin}} & = & {\bf -p}^4/8c^2\\
	{\bf H_{D}}& = & div {\bf E'}(x,z,t)/8c^2\\
        {\bf H_{so}} & = & i {\bf \sigma} \cdot curl {\bf E'}/{8c^2} 
		+ {\bf \sigma} \cdot {\bf E'} \times {\bf p}/{4c^2}.\\
	\end{array}
 \end{equation}

Here ${\bf H_0}$ denotes the standard nonrelativistic Hamiltonian  
in Schr\"odinger form, where ${\bf p}=(p_x,0,p_z)=
(-i\partial/\partial x,0,-i\partial/\partial z)$ is the two-dimensional
momentum operator and  ${\bf A}(z,t)$ is the time-spatial dependent vector   
potential of the  laser field ${\bf E}(z,t)$, which is linearly polarized along the $x$-axis
and propagates in $z$-direction. For the vector potential we include 
the magnetic field component and do not apply the dipole  
approximation, urging us to   perform a two-dimensional  
numerical integration in the $x-z$ plane.
Since ${\bf H_0}$ all other Hamiltonians are $2\times 2$ matrices, they need be considered as 
multiplied by the unity matrix {\bf I} even if not shown explicitely throughout this paper.  
We consider multiply charged ions in the single active electron approximation \cite{sea} which are 
preionized by several or more than ten electrons and thus are easily availabe 
today via lasers \cite{mions} or with highest accuracy via shooting the atoms through  
thin foiles \cite{gsi}.  
Those are well described by the soft-core potential 
\cite{soft} to model the Coulomb field experienced by the active electron of a 
multiply charged ion,  {\sl i.e.} 

\begin{equation}
V(x,z)=- k/\sqrt{q_e+x^2 +z^2}.
\end{equation}

The parameter $k$ is a function of the effective number of
positive charges $Z$ as sensed by the electron. $q_e$ compensates for  
the effect of possible inner electrons and reduced distances of the
electronic wavepacket to the ionic core in two- rather three-dimensional 
calculations.
$k$ may be adapted such that we obtain the correct ionization energy 
for the system of interest with effective charge of the ionic core $Z$ 
and charge of the ion $Z-1$.  
The static field of the ionic core is  expressed by the gradient of the
potential $-\nabla V(x,z)$ and  ${\bf E'}(x,z,t)$ stands for the sum of
this field plus the laser field ${\bf E}(z,t)$.
The following term ${\bf H_P}$ in Eq.(1) indicates the coupling of
the laser magnetic field ${\bf B}$ to the electronic spin as described by the
Pauli matrix $\sigma$. 
The sum ${\bf H_0 +H_P}$ leads to the Hamiltonian in the well known Pauli
equation.  
Further in Eq.\ (1) ${\bf H_{kin}}$ denotes the leading term for the
relativistic mass increase, and ${\bf H_{D}}$ is the well-known
Darwin term. Finally the last term  in the Hamiltonian 
  ${\bf H_{so}}$  stands for the spin-orbit coupling.
Considering our central potential $V(x,z)$ the first term of ${\bf H_{so}}$ in
Eq. (1) disappears because $ \nabla \times (- \nabla V(x,z)) =0$ and the contribution due
to the laser field is of order $1/c^3$. 
Thus, the spin-orbit coupling term becomes  
  
\begin{equation}
{\bf H_{so}} =  {\bf \sigma} \cdot {\bf E'} \times {\bf p} /{4c^2}
%	= f(x,z) {\bf \sigma} \cdot {\bf r} \times {\bf p} 
	=  {\bf \sigma} \cdot {\bf E} \times {\bf p} /{4c^2} + f(x,z) {\bf \sigma} \cdot {\bf L} 
\end{equation}
 
with $f(x,z) = -k(q_e+x^2+z^2)^{-3/2}/{4c^2}$ and where 
${\bf L} =  {\bf r} \times {\bf p}= (0,{zp_x -xp_z},0)$
is the orbital angular momentum, of which 
only the component along the y-direction is non-zero.
The origin of spin-orbit coupling can alternatively be viewed also as being
due to the interaction between the magnetic moment of the electron and the
magnetic field ${\bf B'}$  due to the motion of the positively charged core as
sensed by the electron in its own rest frame.

\subsection{Dynamics and Numerical Approach}
 
 We investigate the dynamics of multiply charged ions exposed to an intense laser field through solving the
following dynamical equation, involving the previously derived Hamiltonian ${\bf H_{FW}}$ in Eq. (1)
(for the convenience, we use the usual atomic units throughout this paper \cite{au}):

\begin{equation}
i\frac{\partial}{\partial t} 
\left ( 
 \begin{array}{c} 
\Psi_{up} (x,z,t) \\
\Psi_{down} (x,z,t) \\
\end{array} \right )
 =  {\bf H_{FW}} 
\left ( \begin{array}{c}
\Psi_{up} (x,z,t) \\
\Psi_{down} (x,z,t) \\
\end{array}\right ).
\end{equation} 

The wave function has two components corresponding to spin-up and spin-down
polarization of the electron and ${\bf H_{FW}}$ is consequently a $2\times 2$ matrix operator. 
The coupling to negative energy states as included in conventional Dirac theory 
is negligible in second order in $v/c$.
The laser field is assumed to be linearly polarized along the
$x$-axis so that the vector potential ${\bf A}(z,t)$ of the laser field 
may be of the form  
   \begin{equation}
 {\bf A}(z,t)= (A_x (z,t), 0, 0) 
   \end{equation}
where the $z$-dependence of the vector potential reflects the propagation of the 
laser pulse in the $z$ direction. 
The spatial dependence of the vector potential indicates that the magnetic 
component of the laser field  $B=\nabla \times {\bf A}(z,t)/c \neq 0$ is 
included and we do not (and can not) carry out the dipole approximation. 
We choose the  vector potential $A_x(z,t)$ to be 
\begin{equation}
 A_x (z,t)= \left \{ \begin{array}{lr} - \frac{cE_0}{\omega
t_{on}} \left [ (t-z/c)sin(\omega t -  \omega z/c) 
 + \frac{1}{\omega}cos (\omega t
-\omega z /c ) \right ] & 0<t-z/c \le t_{on}\\
- \frac{cE_0}{\omega} sin (\omega t
-\omega z /c) & t_{on} <t-z/c< t_p\\ \end{array} \right. 
 \end{equation}
which is associated with a linearly polarized laser field with electric field 
$E_x$ and magnetic field $B_y$ components
\begin{equation} 
 E_x (z,t)= B_y (z,t) = \left \{
\begin{array}{lr} E_0 \frac{t-z/c}{t_{on}} cos(\omega t -\omega z/c) 
& 0<t-z/c \le t_{on}\\ 
E_0 cos (\omega t -\omega z /c) & t_{on} < t-z/c < t_p\\ \end{array} \right.
  \end{equation}

being oriented in the $x$ and $y$ direction, respectively and propagating both in phase in the
$z$ direction. Here, $E_0$ and $\omega$ are the maximal amplitudes of both fields and the 
angular frequency of the laser field, respectively.  Further $t_{on}$ is associated with the linear rising 
time of the laser pulse, {\sl i.e.} 0 is the beginning of the turn-on and $t_{on}$ the end of it.
We note that the vector potential  Eq. (6) may not be continuous at the end of the turn-on phase, however
it is when $t_{on}$ is chosen such that $\omega (t_{on} -  z/c)  =(m+0.25) 2 \pi$ with m being an 
arbitrary integer. The measurable electric field strength however is always continuous.
After the turn-on phase the pulse is assumed to have a constant amplitude till time $t_p$. Obviously a realistic
pulse will turn off afterwards smoothly, however for all observables of interest here this phase is of no interest
and numerical calculations usually terminate at $t_p$.›

Since the laser field is linearly polarized, the interaction term 
involves a term of the form $\bf{p} \cdot {\bf{A}}(z,t)
\rm /c = p_x A_x(z,t) /c$.  which means no coupling between 
momentum and coordinate space. This is because the term couples only 
the $x$ component of the momentum with a function which is dependent
on $z$ but not on $x$. 
Thus, we can still apply the usual split-operator algorithm
\cite{split} to solve the two-dimensional time-dependent 
matrix equation (4) via

 	\begin{equation}
	\begin{array}{rcl}
\left (
\begin{array}{c}
\Psi_{up} (x,z,t+\Delta t) \\
\Psi_{down} (x,z,t+\Delta t ) \\
\end{array}\right )
& = &  exp  [ -i\Delta t (p^2/4-p^4/(16 c^2)) {\bf I} ]\\
~& ~&	  \times exp [-i\Delta t (({\bf p \cdot A}/c +A^2/c^2){\bf I}+{\bf H_D})] 
	    \times exp [-i\Delta t ({\bf H_P+H_{so}}]\\
~& ~&	   \times exp  [ -i\Delta t (p^2/4-p^4/(16 c^2)){\bf I}] 
	  \times
\left (
\begin{array}{c}
\Psi_{up} (x,z,t) \\
\Psi_{down} (x,z,t) \\
\end{array}\right ). \\
	\end{array}
	 \end{equation}	

Here, $\Delta t$ denotes the time step and the unit matrix operator is 
described with {\bf I}.
All exponential operators except the term  $exp [-i\Delta t ({\bf H_P+H_{so}}]$
are diagonal and we apply the split evolution operator on the wave function 
with the help of Fourier transforming  between the coordinate representation 
and the momentum representation. Consequently all derivative operator can be transformed
into multiplications with constants. 
For non-diagonal exponential operators, we usually need to diagonize them.
Fortunately, the non-diagonal operator  $exp [-i\Delta t (H_P+H_{so}]$ involved here  
only depends on the specific Pauli matrix $\sigma$, so that we can carry out the Taylor expansion 
for this exponential operator up to the order of interest here. 
Regarding the interaction term $p_x A_x/c$ special care is needed as mentionned above. Here we do 
the Fourier transformation only for the x-coordinate because of the z-coordinate dependence of the 
vector potential $A_x(z,t)$. This however is sufficient and the exponential function of operators of 
interaction terms on the wave function ends up being merely a sequence of Fourier transformations 
and c-number multiplications.
Because of the splitting of the total Hamiltonian in the exponent we introduce an error following
the Baker-Hausdorff formula because the split terms do generally not commute.
The error of this algorithm is of order $(\Delta t )^3$ between every successive time step. 
Thus, a small time step can ensure to get accurate results; and we have not experienced any problem
with numerical convergence in the regime of interest here.

From the numerical point of view we first solve for the eigenstates of the
bound electron in the ionic core potential by using the so-called "spectral" 
method \cite{feit}. 
In fact, we choose a testing wave function without any symmetries to propagate on 
the two-dimensional potential. This way we obtain all symmetric and asymmetric 
eigenstates of the system.  
As an example, figure 2 shows the energy-level structure
of a model hydrogen-like ion $Mg^{11+}$ of which 
the single active electron experiences the nucleus with charges Z=12. 
Choosing k=80.32 and $q_e=1.0$ we get the corresponding ground-state
energy -72au. 
Note that it is not our purpose to present an exact quantitative model of the ionic level 
structure of an ionic system. We model single electron ions with a softcore rather than a Coulomb potential 
and can adapt the ground state energy by choosing $k$ and $q_e$. With the assumption of a smooting around
the ionic core, we deviate from the exact Coulombic potential and the system 
may also be considered as an active electron plus an ionic core where lower shells are filled and 
inactive. In the intensity range where level structures are important we thus can only make
qualitative statements for the dynamics if we want to associate it with a particular realistic
ion. In the tunneling regime however where structure becomes less important and only
the correct ground state is significant we can be quantitative as well. 

The ground-state wave packet of our model ion relates to a symmetric s-state wavepacket, 
and also the second excited-state wave packet (see figure 2b).
The first, third and fourth excited-states are 
asymmetric (especially not s-states) in space. 
This system will be applied to evaluate the relativistic Stark shift in section III(B).
After obtaining the eigenstates $\Phi_n (x,z)$, we may use the above 
split-step operator in Eq.(8) to investigate the evolution of multiply 
charged ions under the irradiation of external intense lasers.
When the laser pulse is turned on, the system is assumed
to evolve from the spin-polarized ground state, that is, 
the initial condition $\Psi_{up} (x,z,t=0)=\Phi_1 (x,z)$ and
$\Psi_{down} (x,z,t=0)=0$.

\subsection{Observables of interest}

\indent

With the knowledge of the time-dependent wavefunction we are in the 
position to calculate the spatial probability distribution
$|\Psi(x,z,t)|^2 = | \Psi_{up} (x,z,t)|^2  + |\Psi_{down} (x,z,t)|^2$ and  
the expectation value of any  observable ${\cal O}$ 
associated with the system via 

%\begin{equation}
%|\psi(x,z,t)|^2 =  \left ( \Psi_{up} (x,z,t), \Psi_{down} (x,z,t) \right )   
%\left (
%\begin{array}{c}
%\Psi_{up} (x,z,t ) \\
%\Psi_{down} (x,z,t ) \\
%\end{array}\right ) \\
 %	 \end{equation}	

%while for the time dependent expectation value of an oberservable   
%${\cal O}(t)$

\begin{equation}
\langle {\cal O}(t) \rangle = \int_{x_{min}}^{x_{max}} {\rm d}x \int_{z_{min}}^{z_{max}} {\rm d}z
 \left ( \Psi_{up} (x,z,t), \Psi_{down} (x,z,t) \right ) {\cal O}(x,z,t)  
\left (
\begin{array}{c}
\Psi_{up} (x,z,t ) \\
\Psi_{down} (x,z,t ) \\
\end{array}\right ). \\
 	 \end{equation}	

Here the integral turns into a sum in our case as the wavefunction is given on an equidistant
grid from $x_{min}$ to $x_{max}$ in the polarization direction and from $z_{min}$ to $z_{max}$
in the propagation direction.

Particular interest in this article is placed on the spatial average values 
$\langle {x}(t) {\bf I}\rangle$ and  $\langle {z}(t) {\bf I}\rangle$ of the wavepacket and because of their 
relevance for the radiation spectrum on the accelerations in the polarization direction 
$a_x(t)=\langle {\stackrel{..}{x}}(t) {\bf I}\rangle$  and in the propagation direction 
$a_z(t)=\langle {\stackrel{..}{z}}(t) {\bf I}\rangle$ via Eq. (9). 
%Here ${\bf I}$ is the two-dimensional unity matrix.

The radiation spectrum is generally given by a rather complex function of the accelerations
and velocities in all spatial directions (see e.g. Eq. 10.7 in the first review in \cite{atoms}). 
For simplicity we here restrict ourselves to an observation direction perpendicular 
to the 2d plane of motion, i.e. the $y$ direction. Also we are mostly interested in the dominating 
part of the radiation spectrum in the weakly relativistic regime which is polarized in the
polarization direction of the laser field. In the far field spectrum this is proportional to the
squared Fourier transform of simply the acceleration $a_x(t)=\langle {\stackrel{..}{x}}(t) {\bf I}\rangle$.
Also for the purpose of studying weakly relativistic signatures of the spectra it is interessing
to study the less intense spectrum which is polarised in the laser propagation direction and
governed by   $a_z(t)=\langle {\stackrel{..}{z}}(t) {\bf I}\rangle$.
  
Under the assumption of including relativistic corrections up to second order in $v/c$ it is not 
sufficient for $a_x(t)$ to consider the gradient of the potential only as in the nonrelativistic regime.
We rather find in the weakly relativistic limit

\begin{equation}
\begin{array}{ccl} 
\stackrel{..}{x} & = & 
\sqrt {1-\stackrel{.}{x}^2/c^2 -\stackrel{.}{z}^2/c^2}
\times \left ( -\frac{\partial V(x,z)}  {\partial x} 
+{{\stackrel{.}{x}} \over {c}} \left ( {{\stackrel{.}{x}} \over {c}} \frac{\partial V(x,z)}  {\partial x}  + 
{{\stackrel{.}{z}} \over {c}} \frac{\partial V(x,z)}   {\partial z} \right ) 
\right ) 
\\
 & = & \left ( 1+ \frac{3} {2c^2}\frac{\partial ^2} {\partial x^2}
+ \frac{3} {2c^2}\frac{\partial ^2} {\partial z^2} + O(1/c^4) \right )
\times \left ( -\frac{\partial V(x,z)} {\partial x} \right ) \nonumber \\
\nonumber \\
\stackrel{..}{z} & = & \left ( 1+ \frac{3} {2c^2}\frac{\partial ^2} {\partial x^2}
+ \frac{3} {2c^2}\frac{\partial ^2} {\partial z^2} + O(1/c^4) \right )
\times \left ( -\frac{\partial V(x,z)} {\partial z} \right ) \\
\end{array}
\end{equation}

where for the transformation to the second part of the equation we substituted the time derivatives 
via ${\rm d}x/dt = dH_{FW}/dp_x$ and ${\rm d}z/dt = dH_{FW}/dp_z$ and will neglect the higher order terms 
in $ O(1/c^4) $. We note that the first term in the right-hand side is just the non-relativistic limit, 
and the following two terms correspond to  weakly relativistic corrections which will be retained for the 
calculation of spectra. For the evaluation of the spectrum of interest the operators in Eq. (10) need 
just be inserted in Eq. (9) and then to be Fourier transformed. 

%\begin{equation}
%a_x (t)  = \int_{x_{min}}^{x_{max}} {\rm d}x \int_{z_{min}}^{z_{max}} {\rm d}z 
%\left ( \Psi_{up} (x,z,t), \Psi_{down} (x,z,t) \right )  \stackrel{..}{\bf x}  
%\left (
%\begin{array}{c}
%\Psi_{up} (x,z,t ) \\
%\Psi_{down} (x,z,t ) \\
%\end{array}\right ) \\
% 	 \end{equation}	

We now turn to the technical aspects for the evaluation of the photoelectron spectra.
Since the expected ionization is equivalent to a proportion of the electronic wave function 
leaving the vicinity of the ionic core and propagating outwards towards the (unphysical)
 boundaries of the numerical grid, we must avoid reflections of the wave function at those 
boundaries.  This is achieved by absorbing all parts of the wave function approaching the
boundaries by a $cos^{\frac{1}{8}}$ mask function \cite{mask}, which results in a
decrease of the norm of the wave function within the spatial box. 
While quite often  only the part in the vicinity of the ionic core interests, it is here
the opposite as we care only for the part of the wavefunction approaching the absorbing
boundaries.
Assume that we have calculated the wavepacket dynamics up to time $t_f$ in our finite box.
Then we will have information about the photoelectron spectrum from what has been absorped 
till this time at the boundaries at all intermediate times $t_{\alpha}$ and from the ionized 
part of the wavefunction still in the box.  
We begin by treating the absorbed electron flux, say $\Psi_{flux} (x,z,t_{\alpha})$ at time
$t_{\alpha}$ after entering the area of the boundaries. As this part will certainly not
be influenced by the ionic core we will assume this to propagate to the end of interaction 
as a free particle (for details see \cite{apl}). 

At the end of the interaction, at $t_f$, we also have to include the ionized part of the
final spatial wave function still in the box. In order to obtain this from the full wave 
function it will be modified as follows, for example for the spin-up part of the wave function 

\begin{equation}
 \Psi_{out} (x,z,t_f) =  \left  \{  
\begin{array}{lc}
  0 &  \left | x \right | < X_I \\
sin^2 (\pi (\left | x \right | - X_I )/2X_0 ) \Psi_{up} (x,z,t_f)
 & \hspace{1cm}  X_I \le \left | x \right | \le X_I + X_0 \\
 \Psi_{up} (x,z,t_f) & \left | x \right | > X_I + X_0 \\
    \end{array} \right .
\end{equation}

Here $X_I$ represents a range which may be related to the ionic radius 
within which the electron may be considered bound and we will ignore this part of the 
wavepacket as we are only interested in the ionized part of the wave packet.
We call $X_0$  the sliding range for the final wavefunction, where with
increasing distance from the ionic core the wavefunction will 
contribute more to the photoelectron spectrum. Everything beyond $X_I + X_0$ 
can be considered as fully ionized and will fully contribute.
The momentum wave function of the ionized electron can then be obtained by 
fast Fourier transforming both the spatial wave function 
$\Psi_{out} (x,z,t_f)$ and the freely propagated $\Psi_{flux} (x,z,t_{\alpha})$ with 
respect to x.
We note that the main part of the electron distribution is ejected along the 
polarization direction in our range of parameters, so that we focus here on the 
photoelectron spectrum in this direction even though it is analogous in the z-direction.
The momentum wave function describing the emitted electron distribution in x direction 
can be thus expressed as

\begin{equation}
\begin{array}{ccl}
 \Psi_{p} (p_x,z,t_f) & = &  FFT_x \left 
 [ \Psi_{out} (x,z,t_f)\right ] \\
 
  &  & + \sum _{t_{\alpha}} e^{-i p_x ^2 (t_f - t_{\alpha})/2}
FFT_x \left  [ \Psi_{flux} (x,z,t_{\alpha})\right ]. \\  
    \end{array} 
\end{equation}

Here the symbol '$FFT_x$' means fast Fourier transform of the 
wave function with respect to the x-coordinate, and
$p_x$ denotes the electron momentum along the x-axis.  
The summation in the above equation is for all times
$t_{\alpha}$ during which the absorped part of the wave packet is detected. 
The kinetic energy spectrum of the photoelectron can then be obtained by 
integrating the momentum wave function over the z-coordinate,
i.e.  $ P(\epsilon_x, t_f) \simeq (1+\epsilon_x/c^2)/\sqrt{2\epsilon_x} \left | 
\int \Psi_{p} (p_x,z,t_f) dz \right |^2$. 
The prefactor has been derived via ${\rm d}p_x/{\rm d}e_x$ where the kinetic energy 
of the photoelectrons has been approximated via $\epsilon_x \simeq p_x ^2 /2 - p_x ^4 /8c^2$ 
in atomic units, including only weakly relativistic corrections in second order. The 
same procedure is carried out for the spin-down wave function and to obtain the total 
photoelectron spectrum, we sum over both polarisations.

\section {Results and Discussions}

In this section we present and discuss the role of the magnetic laser component (A)
and of the relativistic mass effects (B) on dynamics and radiation of laser driven ions.
In (C) spin effects are investigated regarding the Pauli $\bf \sigma B$ term that merely 
induces spin oscillations and the second order $\bf \sigma L $ term which is responsible 
for spin induced forces. In (D) harmonic generation is studied in the weakly relativistic
regime and in (E) highly energetic above threshold ionization. 

\subsection{Magnetic field effects}

For neutral atoms interacting with moderately intense laser pulses, 
the magnetic component of the laser field can usually be ignored 
in the evaluation of the wavepacket dynamics. Once however the 
velocity of the electron wavepacket becomes nonnegligible as compared to the
velocity of light the Lorentz force $({\bf v}/c) \times {\bf B}$ may
not be ignored as compared to the electric field force of the laser field.
This generally is the case when the relativity parameter $r= e E / (m \omega)$
is not neglible as compared to $c$, i.e. at least $1\%$ depending on the observable 
of interest. 

The magnetic field component of a linearly polarized laser field induces a drift in the laser propagation 
direction \cite{free} which may be strong enough to induce instantaneous ionization even for high laser 
frequencies \cite{stabrel}. There is a regime however where the laser field and correspondingly
the drift is weak enough such that the attraction of singly charged ions can still compensate 
for it and induce recollisions along the propagation direction \cite{David}. If one does not want 
to be restricted in the laser field intensity, a more elegant solution to overcome the drift and avoid
strong ionization would be to employ multiply charged ions. 

In the following we focus on the magnetic field (B-field) induced time evolution of 
the electronic wavepacket in the vicinity of and away from the ionic core, and since
second order relativistic effects turn out to be 
small for the employed laser and ion parameters, the Hamiltonian $H_{FW}$ 
need only include the term $H_0$. Thus, 
we numerically solve the two-dimensional time-dependent Schr\"odinger
equation but beyond the usual dipole approximation and including the 
laser magnetic field, and the vector potential $A=A(z,t)$ consequently 
depends on both t and z. 

In fig.3 we have displayed the dynamics of single electron ions with charge $Z=3$ in a) 
and $Z=4$ in b) in a laser field with a wavelength of 248nm and an intensity 
$1.2\times 10^{17}W/cm^2$ during the pulse duration of 10-cycle constant amplitude.  
We have merely displayed the dynamics of the center of mass of the wavepacket $(<x(t)>,<z(t)>)$ 
in order to compare with our classical intuition. 
For $Z=1$ and $Z=2$ we would have almost complete ionization for the laser intensity 
applied and the figure would ressemble that of a classical free electron drifting with its 
usual ``zick-zack'' in the laser propagation direction \cite{free}. For $Z=3$ we see in Fig. 3a) 
that the ionic core starts to seriously compete with the drift imposed by the laser field,
i.e. there is substantial ionization in the laser polarization and 
propagation direction however also a significant part can still be kept close to the ionic core. 
Interesting to see is that parts of the wavepacket may move opposite to
the laser propagation direction similar to the magnetic recollisions advocated earlier 
\cite{David}. 
We have also considered the center of mass motion restricted to a smaller area around the
vicinity of the nucleus and found dynamics ressembling the ``figure of 8 motion'' known in the 
frame where the drift is transformed away. Simulations of the full wave packet have confirmed
the dynamics towards and away from the nucleus in the laser propagation direction and the
significant amount of ionization.

%We note that here a very large part of the wavefunction ionizes and that we
%have plotted only the center of mass of the part of the wavepacket which has remained in the 
%vicinity of the ion, i.e. in our case within the absorbing boundaries of the numerical box.

%If we further increase the ionic charge, still leaving the laser parameters constant, we find 
%in Fig 3b) again a different behaviour, however stress at this point only, that ionization is 
%now quite small. 
%In order to have a clearer picture how ionization occurs for $Z=3$ ions we have described in Fig 3c)
%also the ionization dynamics with the full wavefunction. Note the dynamics after half a cycle in 
%and against the laser propagation direction and after two cycles of the turn-on clear ionization 
%both in the laser polarization and the laser propagation direction.
%Thus for $Z=3$ ions the ion charge is just too small (or the laser field just too large) to
%avoid substantial ionization. 

A different situation occurs in Fig 3b) for $Z=4$ when the ion charge is high enough 
to avoid substantial ionization and to allow for weakly relativistic bound dynamics. 
The electronic wave packet will oscillate around the ionic core with a particularly 
high velocity in its close vicinity.
 At the times when the electronic wave packet approaches the ionic core, 
the magnetic field component also turns out to be large. Therefore an extremely 
strong Lorentz push of the wave packet around the ionic core in the laser 
propagation direction arises and results in a low electron expectation value 
close to the ionic core \cite{hole}.  
 The "open diamonds" represent the case without magnetic field, i.e.\  
the situation within the dipole approximation. Here the electronic wave packet does 
merely carry out one-dimensional oscillations along the laser polarization 
direction, i.e.\ along the x-axis, as expected. However, when the magnetic component 
of the laser field is included in the calculation, the electronic wave packet is pushed 
towards the laser propagation direction (z-axis) as visible from the full circles. 
This magnetically induced pressure on the electronic wave packet is significantly 
increased around the ionic core as seen in the figure 3b). It results in a low electron 
probability in the vicinity of the ionic core. This can be interpreted as a hole around the
nucleus at (x=0,z=0) as indicated by the arrow in figure 3b). We need emphasize
that we do not put forward holes from knocking out complete inner shell
electrons from multi-electron systems. We rather discuss the strong reduction
of the wave packet expectation value of single active electrons near the
nucleus. For a multi-electron system, however, inner-shell electrons are
likely to be more affected by the magnetically and ionic core induced Lorentz
push investigated here, simply because the attraction from the nucleus is
even larger then. 

It may be obvious that the electron velocity near the nucleus is high
for multiply charged ions. However, we need explain why the
magnetic field, also necessary for the Lorentz force, is significant
at the time of the return of the electronic wave packet to the nucleus.  
When the electronic wave packet returns to the nucleus with essentially 
maximum velocity, the amplitude of the magnetic component of the laser 
field should become nearly zero. This is because velocity and force are 
generally phase shifted by $\pi/2$. However, due to the 
existence of the multiply charged ionic core and the increasing field intensity
during the turn-on of the laser field, an important phase lag can be developed 
during the laser pulse turn-on, which is very critical for the hole formation
 around the ionic core. During the laser pulse turn-on the electronic wave packet 
does not move like a quasi-free particle because of the tight binding of the ionic 
core. When the electronic wave packet is dragged outwards by the laser force, the 
strong ionic core tries to attract it, which results in an important phase lag 
between the wave packet motion and the laser field.  Usually,
when the wave packet returns to the nucleus, this effect is again compensated
because this time the ionic core accelerates the electric wave packet towards
it. However, because of the increasing field in the turn-on phase of the laser 
pulse, the laser force is stronger then,
the Coulomb force comparably weaker and the phase lag consequently only
partially compensated. Once the pulse is at full intensity, the phase lag
remains almost unchanged till the field maximal amplitude begins to decrease
again.

In the following we investigate the electronic relaxation dynamics after the 
laser pulse arising from the reduced circular motion around the ionic core.
For this purpose, we allow the ``hollow'' ion to evolve freely for a time period 
of thirty laser cycles after the laser-ion interaction and during this time 
observe its relaxation dynamics and radiation. We note that  
the hollowing out of the ion due to the magnetic field will be reversed step by
step in space during the relaxation process. In figure 4.a we have shown the
relaxation radiation accompanying this process for the same potential and
laser field parameters as in figure 3b). Apparently, the reverse process to 
magnetically induced hollowing out of the ion gives rise to x-ray emission 
with distinct peaks up to $\sim 230eV$ photon energy. The main peak corresponds 
to the transition between the Stark-shifted first excited state $|1e>$ and the 
ground state $|g>$, as indicated in figure 4.b. Further the 122.67eV peak is 
related to the transition  $|4e> \rightarrow |g>$. Finally we note, that 
the hollowing out of the ion may be optimized, by adjusting the interaction 
parameters such that the Lorentz force ${\bf v}/c \times {\bf B}$ becomes maximal near the 
nucleus.

\subsection{Relativistic Stark shift}

In this subsection, we investigate the role of relativistic corrections to the
Stark shift as visible in the positions of the energy levels of intense laser driven 
ions. For this situation we choose the Hamiltonian
 $H_{FW}=H_0 +H_P+H_{kin}$. We find that the Pauli term has no notable effect on the 
shift of energy levels but we include it for the sake of inclusion of all  first order 
relativistic effects. The leading second order term in the weakly relativistic 
regime is generally $H_{kin}$ which may be associated with a relativistic mass-increase.
We note that the Darwin term is insignificant in its effect compared to the mass increase and 
in fact also to spin-orbit coupling. Since in this subsection we are not interested in 
line splitting  but in the total shift of spectral components, we do not include 
$H_{so}$ here. 
The effect of the mass increase in intense laser driven atoms has been discussed before 
and associated with an enhancement of stabilization \cite{marcos} and with Doppler 
shifts of harmonics \cite{doppler}. 
Here we focus however on the multiphoton regime and how the ionic bound energy levels are 
shifted in addition to the conventional Stark shift. 

We choose the model Z=12 ion as described in section II(B) as target.
The laser parameters involved are the intensity $7\times 10^{16} W/cm^2$,
the wavelength 527nm, and a pulse including a 5.25-cycle turn-on phase and 100-cycle with 
constant amplitude. The whole radiation spectrum is displayed in figure 5, in a) without
including  $H_{kin}$ in the full Hamiltonian and in (b) with including it in the solution 
of the dynamical equation. The relativistic signatures are small on this scale, however, it is 
interesting to note that X-rays via hundreds of photons of the applied laser frequency may be 
generated due to resonances in the multiphoton regime.

We note further numerous resonant structures due to multiphoton transitions among the ionic bound levels in addition 
to smaller peaks displaced by up to few photon energies of the applied laser field to the resonant lines. 
In order to have a clearer picture of the structure and the deviation due to the relativistic mass shift, 
we have depicted the enlargements of the first three dominant resonant spectral structures in 
  figures 6, 7 and 8, respectively for the multiphoton resonances of the ground state $|g>$ to the 
first excited state $1e>$, to the second excited state $|2e>$ and to the fourth excited state $4e>$. 
The third excited state $|3e>$ resembles a d-state and we find that the resonance on $|3e> \leftrightarrow |g>$ 
involves a coupling three orders of magnitude smaller than those of $|1e> \leftrightarrow |g>$ and 
$|4e>\leftrightarrow |g>$. We emphasize that our calculations are beyond the dipole approximation so that
the usual selection rules do not apply. In these three figures the solid line and dashed line
represent the cases in which relativistic corrections due to the mass shift are ignored and included, 
respectively.

In figure 6, we display the dominant resonant line which is
located at $87.88\omega$ following the calculation without the relativistic 
mass shift correction. This corresponds to the energy difference between 
the ground state and the first excited state including the conventional 
nonrelativistic Stark shift. 
When the leading relativistic correction to the electron kinetic energy is 
taken into account, we find from the dashed line in Fig 6 the corresponding 
resonance shifts to $87.45\omega$, with $\omega$ denoting the laser frequency. 
The additional redshift of the resonance transition of $ |1e>\leftrightarrow |g> $ 
is as high as $0.43\omega$ for the parameters chosen here. We refer to the correction
 as  the "relativistic Stark shift". 
Furthermore, we note clear peaks displaced by two laser photons to both sides of the 
main resonanant lines; i.e. the laser stimulates the absorption or emission of two photons
prior the emission of the high frequency photon from the resonant state. The effect of the
relativistic Stark shift is clearly visible from the comparison of the dashed and solid line.
In addition, small lines are visible which are displaced at about one-photon away from the resonances.
Those would be forbidden under the dipole approximation, however appear because of quadric and high order 
contributions included in our calculation.
Those are higher order terms and small in the weakly relativistic regime of interest here.
Thus, higher order corrections to the position of this small peak as by the relativistic Stark shift 
are even smaller and are not visible for this peak displayed by one photon around the $ |1e>\leftrightarrow |g> 
$ transition. However, we note here already that it will be visible for higher excited states, 
for example for the $|4e>\leftrightarrow |g>$ transition in figure 8.   

For the resonance between $|2e> \leftrightarrow |g>$, the result is plotted 
in figure 7. It is similar to figure 6, but the relativistic Stark shift is only $ 
0.39 \omega$. As addressed in section II(B), the second excited state $|2e>$ is symmetric 
in space and ressembles an s-state similar to the ground state $|g>$,
so that the resonant transition $|2e> \leftrightarrow |g>$ is considered. 
Thus only  high order terms beyond the dipole term contribute. Considering the scale 
of the vertical axis, we find that the resonant signal is three orders of magnitude smaller 
than in the case for the  $|1e> \leftrightarrow |g>$ transition.
In addition the relativistic Stark redshift turns out smaller than in the case of 
the $|1e> \leftrightarrow |g>$ resonance, although the $|2e>$ state is 
not so tightly bound as the state $|1e>$. Moreover, the one-photon
spaced peaks are no longer visible for the $|2e> \leftrightarrow |g>$ 
transition.  If we continue to explore the $|4e> \leftrightarrow |g>$ 
resonance, we find that the relativistic Stark redshift becomes relatively 
large again, as indicated in Fig. 8.
Since the state $|4e>$ is asymmetric and resembles a p-state as $|1e>$, the dipole 
transition $|4e> \leftrightarrow |g>$ is permitted. Further the
state $|4e>$ is less strongly bound by the ionic core and consequently 
much easier influenced by the external field. Therefore, the 
relativistic Stark shift attains  $ 0.82 \omega$ and is thus larger than in the 
cases for the transitions $|1e> \leftrightarrow |g>$ and $|2e> \leftrightarrow |g>$.
Finally the peaks spaced one-photon away from the resonance appear again as already 
in figure 6, however the relativistic Stark redshift becomes visible here also for 
these small peaks.

Figure 9 shows how the energy levels of the multiply charged ion of interest 
move under the radiation of an intense laser field. In the left part, the  
column (a) represents the situation without laser field and the central part (b)
indicates the inclusion of the usual Stark shift but without relativistic corrections. 
The column (c) includes the  "relativistic Stark shift" in addition to the usual 
Stark shift, which arises from the leading relativistic correction to
the kinetic energy of the electron. 
The detailed position of the  states $|1e>, |2e>$ and $|4e>$ is enlarged in the 
right part of figure 9. 
The series of transitions in columns (b) and (c) are corresponding  
to the solid lines and dotted lines in figures 6, 7 and 8, respectively.
This figure gives us a clear picture of how the energy levels shift under an 
external intense laser field.
The relativistic Stark shift, though small when compared to the absolute value 
of the energy levels, can clearly be identified in the radiation 
spectra, and should in principle be measurable in experiments. 
 
\subsection{Spin effects}

We now turn to the role of the spin degree of freedom of the electron in intense laser-ion interaction. 
In first order in $v/c$ the Pauli equation already includes the coupling of the magnetic laser field to the 
spin as noted earlier \cite{spineffect}. We will not discuss this here and are more interested in the interaction 
of the spin degree of freedom with an operator describing the electron rather than the laser field. This is 
given by spin-orbit coupling arising first in second order in $v/c$.

With the laser intensity and ion charge increasing, such that second order effects become important, spin 
signatures in the dynamics are expected to be visible as noted first recently \cite{spin}.
 For free electrons, similar spin-induced forces have also been noted almost purely analyticaly, especially 
most interestingly a small one in the direction of the magnetic field of the laser field\cite{spinmarkus}. 
Here, we concentrate on the effect of the spin degree of freedom on bound electron dynamics and radiation 
in intense laser fields. 
In this case the used Hamiltonian $H_{FW}$ includes all second order terms in $1/c^2$ as shown in equation (1), 
of which both the Pauli term $H_P$ and the spin-orbit coupling term contribute to spin effects. 
As noted in subsection A the Lorentz force may induce a significant angular motion around the ionic core 
with considerable orbital angular momentum $\bf L$ with respect 
to the origin set by the nucleus.  
 We show that the resulting enhanced spin orbit coupling gives rise to 
observable effects in the electron dynamics and radiation. In particular we 
note a significant splitting of the non-symmetric bound states due to this 
additional interaction which leads to well separated doublets and four-line  
structures in the radiation spectra.  
In general terms we understand those also as indications that  
the influence of the spin and other relativistic effects  
are both principally observable in experiments and nonnegligible in theoretical  
treatments for relatively low laser intensities. 
 
We are still interested in the weakly relativistic regime of optical laser intensities 
of up to at about $ 10^{17}$Wcm$^{-2}$, which have been implemented in several laboratories 
worldwide and which still allows for laser - bound electron dynamics. 
We employ the 527nm (double Nd:glass) laser with an intensity of $7\times 10^{16}W/cm^2$
 to interact with Z=12 ions, which are intially spin-up polarized. 
The laser pulse has a 5.25-cycle turn-on and 100-cycle constant amplitude duration.  
In figure 10 we address the spin polarization itself and have displayed  
the expectation value of the electronic 
wavefunction in ``spin-down'' configuration as a function of the interaction
time.  We compare results from the full second order Hamiltonian ${\bf H_{FW}}$ including 
spin-orbit coupling with those where spin-orbit coupling 
${\bf H_{so}}$ has been ignored. Both situations involve a 
spin flipping with twice the laser frequency. With spin-orbit coupling however
the total flipping amplitude is higher because of a second oscillation due
to the magnetic field of the frame of reference transformed nucleus ${\bf B'}$.
 Finally the figure shows an effective polarization due to spin-orbit coupling 
in the turn-on phase, while without spin orbit coupling the electron periodically 
returns to the initial polarization in complete spin-up configuration. 
 %Such an effective spin polarization means that the electron senses a magnetic 
%field ${\bf B'}$ in its own rest frame in the case where spin-orbit coupling is involved.
 
The most significant qualitative features appear in the radiation
spectrum in terms of strongly laser-enhanced line splitting.
In figure 11 we have displayed the radiation spectrum of light emitted perpendicular 
to the plane spanned by the laser polarization and propagation directions and 
being polarized in $x$-direction.   
The upper row describes the situation governed by the Pauli Hamiltonian 
while the lower involves the full second order Hamiltoninan ${\bf H_{FW}}$ in 
Eq.(1). Figures 11i(a), 11i(b) and 11i(c) show the spectral segments corresponding to
the resonances of the first excited state $|1e>$ to the ground state $|g>$, the 
third excited state $|3e>$ with the ground state $|g>$, and  
the third excited state $|3e>$ to the first excited state $|1e>$. 
Comparing the upper and lower row we note shifts and splittings of the spectral components 
into a doublet in (a) and (b) and a four-line structure in (c).  
In addition, the relativistic Stark redshift is also observable, as discussed at length 
in III(B).
We found also that the Darwin term due to ${\bf H_D}$ has no notable effects in this 
situation. We confirmed that the spectral 
features displayed are generally well separated for ions with different charge states, 
should a possible experiment not allow for a pure sample of the ion of choice. 
 
We explain the doublets and four-line structure with the splitting of the asymmetrical 
excited states $|1e>$ and $|3e>$ due to the additional spin orbit interaction 
as depicted in Fig.\ 11ii , while the symmetrical states, possibly $s$-states, 
remain unchanged.  
 The splitting becomes larger with increasing laser intensity or charge of the  
ionic core.  
All transitions give rise to spectral features because the common selection rules  
do not apply in the parameter regime beyond the dipole approximation as investigated  
here. The bound states in Fig.\ 11ii drawn with thick lines indicate  
that those states of the split doublet are most populated explaining 
the relative heights of the spectral lines in Fig.\ 11i. 
The amount of population in the excited states is well above $1\%$ (e.g. $1.25\%$ in the 
first excited state) such that the total radiation should not be insignificant.
We note that the amount of the splitting is $\frac{\Delta 
E}{\omega_L}  \simeq  0.48  $ for the state $|1e>$ and $0.04 $ for the 
state $|3e>$ (here, $\omega_L=0.0866$ a.u. is the applied laser frequency) so 
that the enhanced spin-orbit splitting should be easily measurable in 
experiments. Comparing those values with the amount of spin-orbit splitting 
without the presence of the laser field we have evaluated numerically via the 
same techniques $\frac{\Delta E_0 }{\omega_L} \simeq  0.042 $ for $|1e>$ and  
$0.004$ for the state $|3e>$. Thus, we find that the total enhancement factor 
of the spin-orbit splitting due to the intense laser field 
for our set of parameters amounts to approximately $\simeq 10$. 
We note that those values should increase considerable for more intense  
laser fields and for higher charged ions. However, we also emphasize that  
for less charged ions, in particular hydrogen, spin orbit coupling has still little significance.

\subsection{Generation of Coherent keV X-rays}

From nonrelativistic laser atom interaction it is well known that parts of the electronic
wavepacket, for appropriate parameters,  may tunnel through the Coulomb barrier, then propagate
essentially freely in the laser field and when returning to the ionic core in the oscillating 
field may release their energy in form of radiation \cite{hhgrule}.
 The corresponding radiation spectrum in terms of harmonics of the applied field is in fact 
quite attractive because quite high frequency coherent light is achievable. Instead of an exponential decay
of the spectrum with respect to the harmonic order we here experience a plateau and a cut-off energy
of the emitted harmonics  at $I_p +3.17U_p$. Here $I_p$ is the ionization potential 
of the ions and $3.17 U_p$ the maximal kinetic energy of the electron at the time of the return to
the nucleus. This law is valid in the tunnel regime where the Keldysh parameter 
need fulfill $\gamma_K = \sqrt{{I_p} \over {2 U_p}} \ll 1$.

The kinetic energy of electrons increases significantly with the laser intensity. Consequently higher 
harmonics are expected with an increasing laser intensity if the ionic potential is raising correspondingly, 
as shown recently already in the nonrelativistic regime with an output of more than 2000 low-frequency harmonics 
\cite{1000HHG}. 
We here concentrate on the weakly relativitic regime and show that the harmonic order increases considerably
by enhancing the charge of the ion and simultaneously increasing the laser field. Here we took special care on
adapting the ratio of $U_p$ and $I_p$ such that we still remain in the tunneling regime. 
  
In our calculations, we employ the $KrF$ laser (wavelength $\sim 248nm$)
to interact with multiply charged ions of charge Z=3 and Z=4. 
There are two advantages to use short-wavelength lasers:
(1) the generated harmonics have a high efficiency and are well 
separated even near the cut-off because of 
 $\omega_{2n+1}-\omega_{2n-1} = 4  \pi /\lambda$ ($\lambda$ being 
 the fundamental laser wavelength); 
(2) the numerical resolution is better, as the box size is proportional 
to the square of laser wavelength and the CPU time to the cube. 
Moreover, the chosen laser wavelength is available in experiments.
The pulse envelope is assumed to have a 10-cycle linear turn-on, followed by
 10-cycles with constant maximal amplitude. For Z=3 ions (potential parameters
k=6.48, $q_e=1.0$), we use the laser intensity of $2.5\times 10^{16} W/cm^2$.
The result is shown in figure 12, in which we find as highest harmonic the 131st order. 
To our best knowledge, the maximum harmonic order ever obtained in experiments for this 
short driving wavelength is the 37th \cite{harmonic3}. There the authors expressed 
that they believe that the $He^+$ ions may have also contributed to the harmonic emission.
Our numerical simulation showns that this harmonic spectrum can be extended to the 
131st order if the $Z=3$ ions ($Li^{2+}$) are employed instead.
This harmonic has a wavelength  $\sim 1.9nm$, and the
efficiency indicated is still of the level $10^{-10} \sim 10^{-11}$ relative to
the radiation at the fundamental frequency.
Please note in figure 12 that we have only shown the radiation polarized in the laser polarization 
direction as the radiation polarized in the laser propagation direction is relatively small 
for the parameters chosen here.

Coherent X-rays even in the keV regime can be obtained  when $Z=4$ ions (potential
parameters k=10.7, $q_e=1.0$) are used. In figure 13, we display the harmonic
spectrum of $Z=4$ ions driven by a laser pulse with the same parameters as in the
previous figure but with an intensity of $ 10^{17} W/cm^2$. 
The spectrum polarized in the laser polarization direction in a) has been evaluated via
the Fourier transform of $a_x(t)$, and the smaller contribution polarized in the laser 
propagation direction in b) via $a_z(t)$ in Eq. (10). The cut-off is enlarged 
as an insert in the right corner of figure 13 a). We emphasize that coherent radiation
of the 427th harmonic is clearly observable in spite of the low efficiency of $10^{-13}$
relative to the radiation at the fundamental frequency.
The photon energy of this harmonic exceeds even 2keV and shall be useful for many applications,
as e.g. for time-resolved X-ray diffraction.
For the parameters chosen we find no significant deviation to the cut-off rule, even though the
factor 3.17 is expected to alter with more than weakly relativitic free electron dynamics 
changing the kinetic energy at the recollision time. Most remarkably, and unfortunately, 
the plateau is tilting with 
increasing charge reducing the efficiency of the highest harmonics substantially. 
Thus our results indicate that a part of the electronic wave packet can still tunnel
through the barrier formed by a deep ionic potential but the amount that tunnels out
and returns reduces drastically with the ion charge. Thus with further increasing
charge higher harmonic energies are possible but special care is devoted to the
small efficiency.

The calculations for figure 13 have been carried involving all first order terms in $v/c$,
i.e. including the role of the magnetic field, and the leading terms in $(v/c)^2$. We have chosen 
parameters in the weakly relativistic regime such that higher terms will not make a difference
in the results presented. We note that the spectrum in figure 13 a) in fact would look hardly 
different without all the terms in $v/c$, however the one in figure 13 b) would hardly ressemble the
correct one depicted. This is not surprising as the laser induced dynamics in the laser propagation
direction is fully induced by the magnetic field component of the laser field.  

Regarding experimental possibilities for high harmonics from multiply charged
ions in the coherent keV X-rays, there are two ways at hand. There may be
the double pulse experiment \cite{doublepulse} where the first pulse is used to 
strip several or many electrons of the neutral atoms and then a  moderately delayed
pulse will interact with the obtained multiply charged ions and generate the 
coherent keV X-rays as described above.
 However, here the free electrons ionized by the first pulse may influence the phase 
matching of harmonic emissions and there may also be a mixture of many charge states.
The alternative way is to shoot atoms through thin foils which is clearly a cleaner 
way to generate ions with a pure charge state \cite{gsi}. 
Extremely high densities of multiply charged ions have been generated recently 
\cite{exhighion}, which raises hope for a reasonable efficiency for ion charges 
well above those investigated here.

\subsection{High-order Above-threshold-ionization}

When laser field strengths are employed which are high with respect to the binding fields of the
ionic core, a large part of the electron wavepacket will escape the vicinity of the ion and not 
return. Those electrons may in fact be quite energetic and will be of interest in this section. 

We begin by calculating the photoelectron spectrum for $Z=3$ ions with the method described in 
section II(C).  Here, we use a laser pulse with a 3-cycle turn-on, and  10-cycles with constant 
maximal amplitude. The applied maximal intensity is equal to $2.5\times 10^{16} W/cm^2$, and the 
laser wavelength is 248nm (corresponding to a photon energy near $ 5eV$). The resulting photoelectron 
spectrum is displayed in figure 14. The spectral range from 860eV to 900eV is enlarged as an insert 
in the right-up corner of this figure. We note very energic electrons with energies up to 2keV with 
the usual photon-energy spacing characteristics of photoelectron peaks. Figure 15 shows the case 
for $Z=4$ ions with an appropriately higher laser intensity of $1.2\times 10^{17} W/cm^2$. 
As expected more energetic (above 5keV) photoelectrons can be deteced. A rather regular spacing of the
peaks in this above-threshold-ionization spectrum is still notable in this keV energy regime. 

Thus next to extremely high harmonics we find also very energetic electrons due to above-threshold ionization.
In more general terms those results and the ones in the previous section show that very high-order nonlinear
effects are governing the interaction of multiply charged ions with very intense laser fields.

\section{Conclusion} 

We believe to have shown that the physics of multiply charged ions in very intense laser fields is even richer than 
that for neutral atoms with moderately intense laser fields. There is not merely the effect of scaling the known effects 
for neutral atoms to the new intensity regime but the upcoming of many relativistic influences imposes a fundamentally 
different dynamics. We have seen that the magnetic field component of the laser field completely modifies the
dynamics and may even induce a partially circular motion around the nucleus with the effect of a reduced expectation value 
just in the near vicinity of the ionic core. In the regime where second order terms in $v/c$ are important, the
relativistic mass shift modifies clearly the positions of the spectral components in the multiphoton regime. 
Amplification was found on 
resonances involving of the order of 100 photons with interesting non-dipole spectral features displaced few 
harmonics away from the atomic resonances. The spin, usually of no importance in the nonrelativistic regime, starts
to oscillate and via spin-orbit coupling modifies substantially dynamics and radiation. The laser enhanced splitting 
of resonant spectral lines is a clear relativistic quantum signature.

Those aspects are associated with fundamentally new influences in the weakly relativistic regime with respect to the
nonrelativistic case and were most conveniently described by applying the expansion of the Dirac equation up to 
the second order in $v/c$. This allowed us to relate each effect to different parts of the Hamiltonian and to carry
out our numerical investigations even for the low frequencies available in most present day high power laser systems.

We stressed also that the combination of highly charged ions and high power lasers can be useful for applications. 
High-order above threshold ionization was shown to give rise to photoelectrons in the multi-keV regime already for 
laser intensities around $10^{17}W/cm^2$. More interestingly parts of those highly energetic electron wavepackets 
may return to the nucleus and we indicated also coherent high harmonics in the keV regime.
Even though for high harmonic generation towards the coherent hard X-ray regime the quantitative aspect
appears most attractive, we pointed out also qualitative changes as the problematic titling of the plateau of the 
harmonic spectrum with increasing ion charge. 

\vspace{1cm}

\indent
The authors acknowledge funding from the German Research Foundation (Nachwuchsgruppe within Sonderforschungsbereich 276). 
SXH would like to thank W. Becker for fruitful discussions and acknowledges present funding from the Alexander 
von Humboldt Foundation. CHK acknowledges helpful discussions with M. Casu, J. Ullrich and M. W. Walser.

\rm \noindent $^*$ {\sl\small Present address: Max-Born-Institut for nonlinear
Optics and Short Pulse Spectroscopy, Rudower Chaussee 6, 12489 Berlin, Germany. Present email: suxinghu@mbi-berlin.de }\

\noindent $^{**}$ {\sl\small Email address:  keitel@physik.uni-freiburg.de }

%\newpage
\begin{figure}
 \caption{Schematic scetch relating near-optical laser intensities to the charge state $Z$ of a single 
 electron ion such that the laser electric field and that of the nucleus become comparable, i.e. the laser 
 field induces a nonperturbative interaction without large ionization. The emphasis of this paper is on the 
weakly relativistic regime from $Z$=3 to $Z$=12} 
 \end{figure}

\begin{figure}
 \caption{2.a) The energy-level structure of a model 
multiply charged ion with Z=12. The potential is described by a soft-core model 
$-k/\sqrt{q_e+x^2 +z^2}$ with k=80.32 and $q_e=1.0$ and ground state energy of -72au..
 2.b) The probability distributions of the electronic wavepackets for our model multiply 
 charged ion for the three lowest energy states $|g>, |1e>$ and $|2e>$. 
} 
 \end{figure}

\begin{figure}
\caption{ 
 The center-of-mass evolution of the electronic wave packet 
for the cases with B-field ("full circles") and without B-field ("open diamonds"). 
The laser parameters involve a wavelength of 248nm, an intensity of $1,2\times 10^{17}W/cm^2$, 
a three-cycle linear turn-on and a ten-cycle constant amplitude duration. 
For $Li^{2+}$ in a) (Z=3, $q_e=1.0, k=6.48$) there is a significant drift in the laser propagation
direction  and in addition to substantial ionization (not visible) a considerable part of the 
wavepacket returns to the nucleus opposite the laser propagation direction. 
A reduction of the expectation value of the electronic wave packet is observed around the close vicinity of
the ionic core for $Be^{3+}$ (Z=4, $q_e$=1, k=10.7) as indicated by the arrow in b),  
when the B-field is considered.   
%In c) snapshots of the wavepacket are depicted for the parameters of
%a). For fig3a, there are only 5 cycles at full strength and the box size 
%is only 102.4au along the polarization direction and propagation direction 6.4. a.u.. 
}
\end{figure} 
 
\begin{figure} 
\caption{ 
a.) The relaxation radiation spectrum  of the hollow ion ($Be^{3+}$ 
as prepared for fig. 3b) during the time of thirty laser cycles after the laser 
pulse. The strongly reduced expectation value close to the nucleus of the 
active electron is reversed again and x-ray radiation is emitted. 
The corresponding transitions are plotted in b), which are modified by the Stark shift.} 
\end{figure} 

\begin{figure} \caption{The spectrum of Z=12 model ions (with level structure indicated by Fig.2), 
which are exposed to an intense laser pulse of intensity 
$7\times 10^{16}W/cm^2$ and the laser wavelength 527nm. 
The laser pulse has a 5.25-cycle turn-on, followed by a 100-cycle period with constant amplitude.  
The laser parameters are in the multiphoton regime with respect to the ion charge so that
resonant structures are visible. 
The solid line and the dotted lines represent the cases without and 
with the leading relativistic correction, respectively.} 
\end{figure}
 
\begin{figure} \caption{The segment of Fig.5 for the multiphoton resonant transition 
between the bound states $|1e> \leftrightarrow |g>$.  
The solid  and dotted lines represent the cases without and 
with the leading relativistic correction, respectively.
A clear redshift of $0.43\omega$ of the resonant line is observed when the leading
relativistic correction is taken into account. It can be attributed 
to the relativistic Stark shift of the energy levels of the ion.} 
\end{figure}

\begin{figure} \caption{
Same as Fig.6 but for the resonant transition $|2e> \leftrightarrow |g>$. 
The redshift is up to $0.39\omega$ for this case.} 
 \end{figure}

\begin{figure} \caption{
Same as Fig.6 but for the resonant transition $|4e> \leftrightarrow |g>$.
A large redshift of $0.82\omega$  for this resonant transition is visible
%, which appears likely  to be measurable.
.} 
 \end{figure}
 
\begin{figure} \caption{ 
Energy level diagram of the four lowest eigenstates of the Z=12 ion in intense 
laser field. 
In the left part, the column (a) shows the eigenstates without external fields, 
column (b) involves the conventional Stark shift and the relativistic Stark shift is included in column (c). 
The transitions in columns (b) and (c) are corresponding to the solid and dotted lines in figures 6, 7 and 8. 
The details of the shifting process are enlarged in the right part for states $|1e>$,$|2e>$ and $|4e>$.  
} 
\end{figure}

\begin{figure} 
\caption{The dynamics of the spin degree of freedom for the situations 
without (a) and with (b) spin-orbit interaction as viewed  
from the population of the wavefunction with spin-down polarization. 
The initial electron is spin-up polarized.  
The intense laser-enhanced spin-orbit coupling  
leads to an effective spin polarization in the  
turn-on phase and an additional oscillation due to 
magnetic field from the nucleus in the rest frame of 
the electron. The laser parameters involve a wavelength of 527nm, an 
intensity of $7\times 10^{16}W/cm^2$, a 5.25-cycle linear turn-on 
and a 10-cycle duration with constant amplitude. 
The parameters for the ionic core are $q_e$=1, k=80.32 (Z=12 ions).}
  \end{figure} 
  
\begin{figure} 
\caption{ 
i). Radiation spectrum of the laser driven Z=12 ion. The 
first row corresponds to the Pauli modelled system, and the second row is for 
the case where spin-orbit coupling including the relativistic mass shift and  
Zitterbewegung is taken into account. Figures (a), (b) and (c) are 
associated, respectively, with transitions from $|1e>$ to $|g>$, $|3e>$ to 
$|g>$,and $|3e>$ to $|1e>$ (see also Fig.11.ii). The spectral lines split into 
doublets ((a) and (b)) and a four-line structure (c)  due to the spin-orbit 
interaction.  All parameters are same as those in figure 10 part from 100 cycles
at full strength rather than 10;\\ ii). The schematic 
diagram of state-splitting as induced  by the enhanced spin-orbit interaction due 
to the intense laser field. We note that asymmetric states split as opposed to  
symmetric states. Transitions (a), (b) and (c) are associated with the corresponding 
spectral lines in figure 11.i).} 
 \end{figure} 
 
\begin{figure} 
\caption{The harmonic spectrum emitted from Z=3 ions with parameters k=6.48 and 
$q_e=1.0$ characterizing the potential. The laser intensity is $2.5\times 10^{16}W/cm^2$, 
and the $KrF$ laser wavelength with 248nm is applied. The laser pulse has a linear 10-cycle 
turn-on and is  followed by a 10-cycle duration with constant amplitude.
The 131st harmonic is visible in the the cut-off frequency. }
 \end{figure}

\begin{figure} \caption{Same as Fig.12 but for Z=4 ions (k=10.7, $q_e=1.0$).
The laser intensity is further increased to $10^{17}W/cm^2$ resulting in harmonic emission 
 up to the 427th order with a photon energy well above 2keV. Consequently
coherent keV X-rays can be obtained through high harmonic generation with multiply 
charged ions. The contribution of the harmonic yield in the laser polarization direction is 
depicted in a) and the one in the laser propagation direction in b).}
 \end{figure}
 
\begin{figure} \caption{The photoelectron spectrum for Z=3 ions.
 The laser parameters are the same as those in Fig.12 but with
 a steeper turn-on of 3 cycles. 
 We can observe very  energetic (2keV) 
photoelectrons and the ATI peaks are spaced by the photon energy even 
at electron energies close to the order of 1 keV.}
 \end{figure}

\begin{figure} \caption{Similar to Fig.13 but for Z=4 ions.
 The laser parameters are the same as those in Fig.14 but involving 
 a the higher intensity of $1.2 \times 10^{17}W/cm^2$. 
More energetic $\sim 5keV$ photoelectrons are ejected, while conventional ATI 
features are still conserved.}
 \end{figure}
 
%******************************************************* 

\newpage

\begin{figure} 
\begin{center} 
\unitlength1cm 
   \makebox {\epsfysize 15cm \epsffile{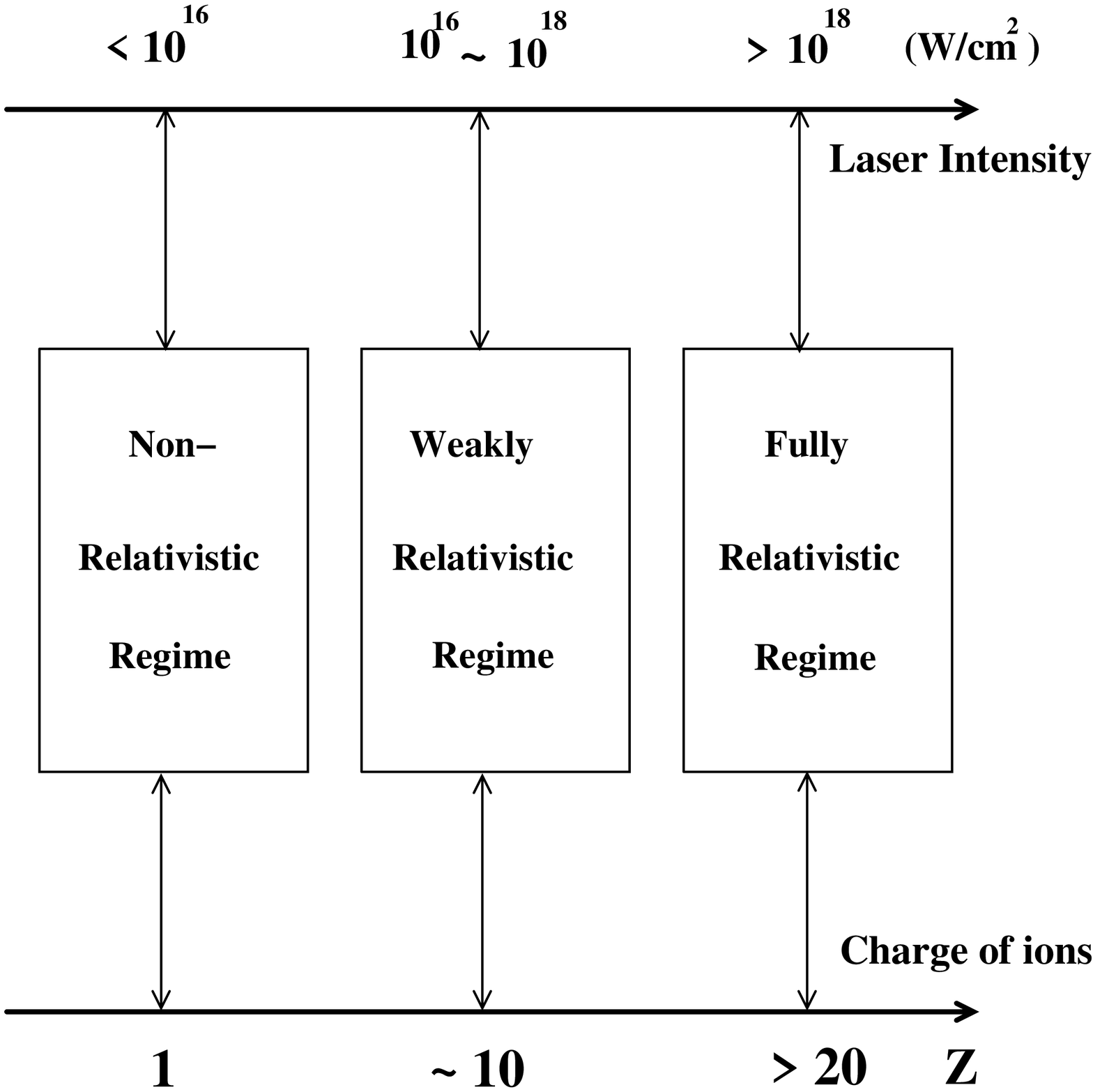}} 
%  \makebox {\epsfysize 12cm \epsffile{hufig01.ps}} 
\end{center} 
\end{figure} 
\vspace{1cm} 
 
Fig. 1: S. X. Hu  and C. H. Keitel, ``Dynamics of ...'' 

\newpage
 
\begin{figure} 
\begin{center} 
\unitlength1cm 
   \makebox {\epsfysize 15cm \epsffile{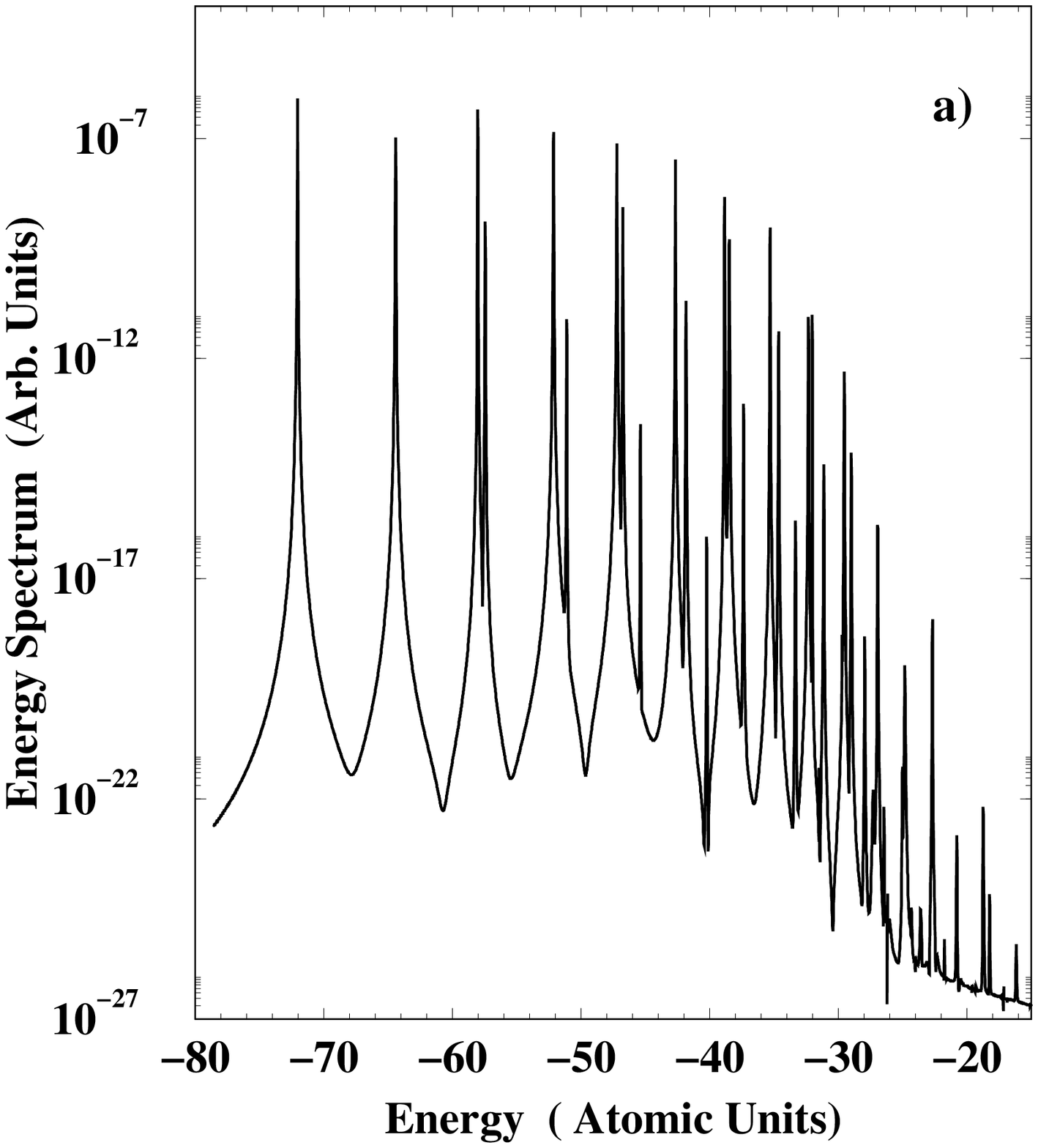}} 
\end{center} 
%\vspace{1cm} 
\end{figure} 
 
Fig. 2a: S. X. Hu  and C. H. Keitel, ``Dynamics of ...'' 

\newpage

\begin{figure} 
\begin{center} 
\unitlength1cm 
   \makebox {\epsfysize 18cm \epsffile{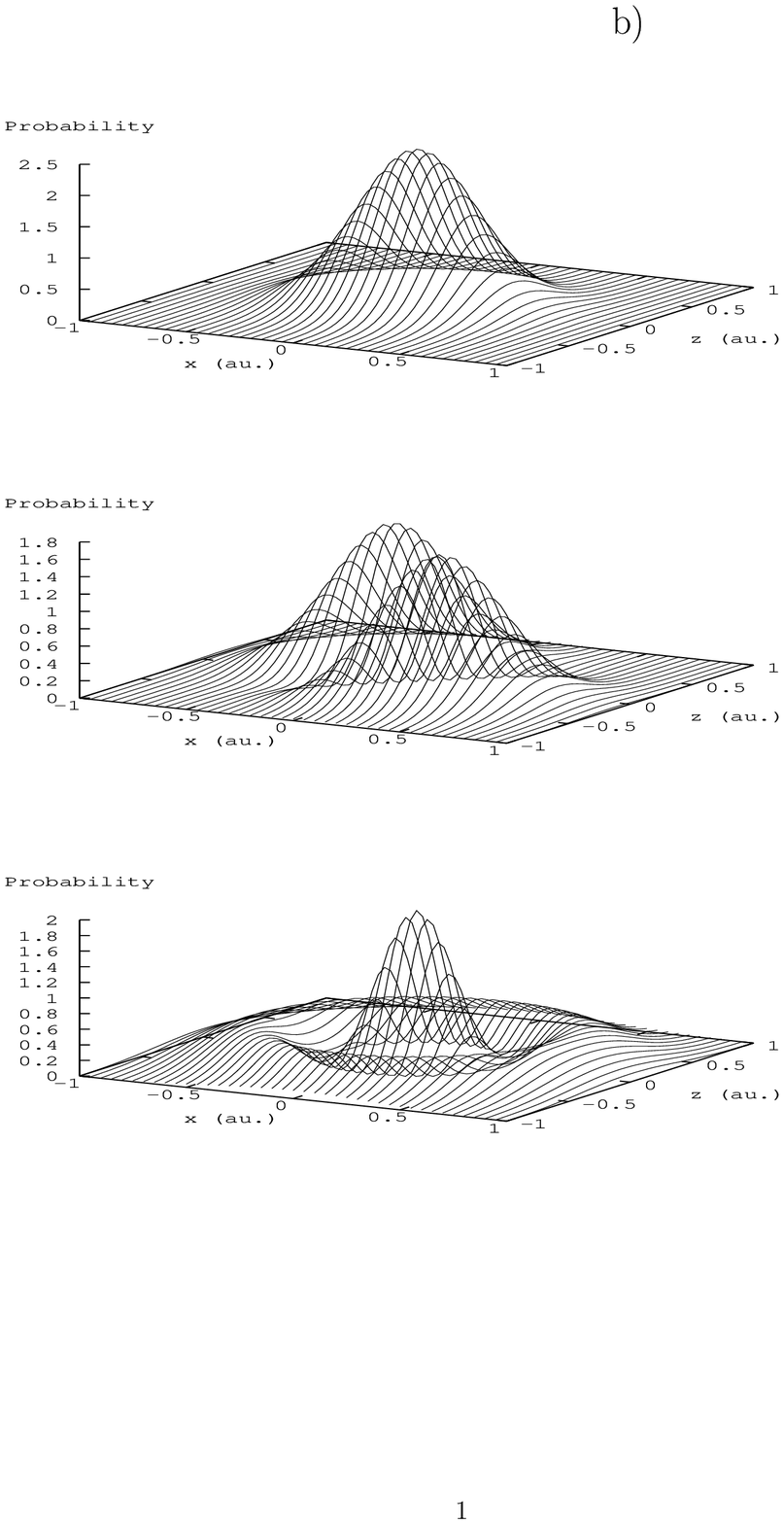}} 
\end{center} 
%\vspace{1cm} 
\end{figure} 
 
Fig. 2b: S. X. Hu  and C. H. Keitel, ``Dynamics of ...'' 

\newpage

\begin{figure} 
\begin{center} 
\unitlength1cm 
   \makebox {\epsfysize 15cm \epsffile{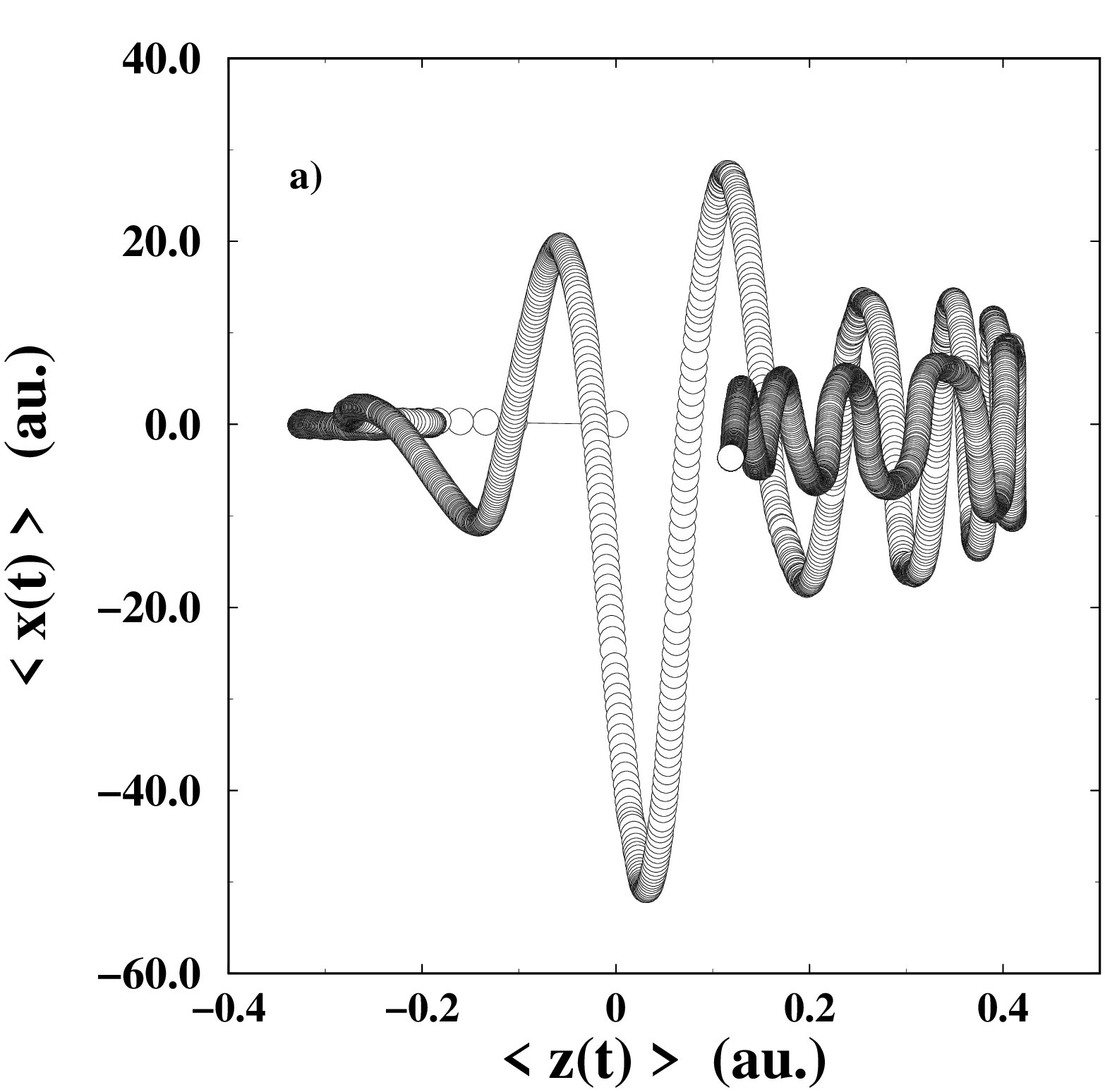}} 
\end{center} 
%\vspace{1cm} 
\end{figure} 
 
Fig. 3a: S. X. Hu  and C. H. Keitel, ``Dynamics of ...'' 

\newpage

\begin{figure} 
\begin{center} 
\unitlength1cm 
   \makebox {\epsfysize 15cm \epsffile{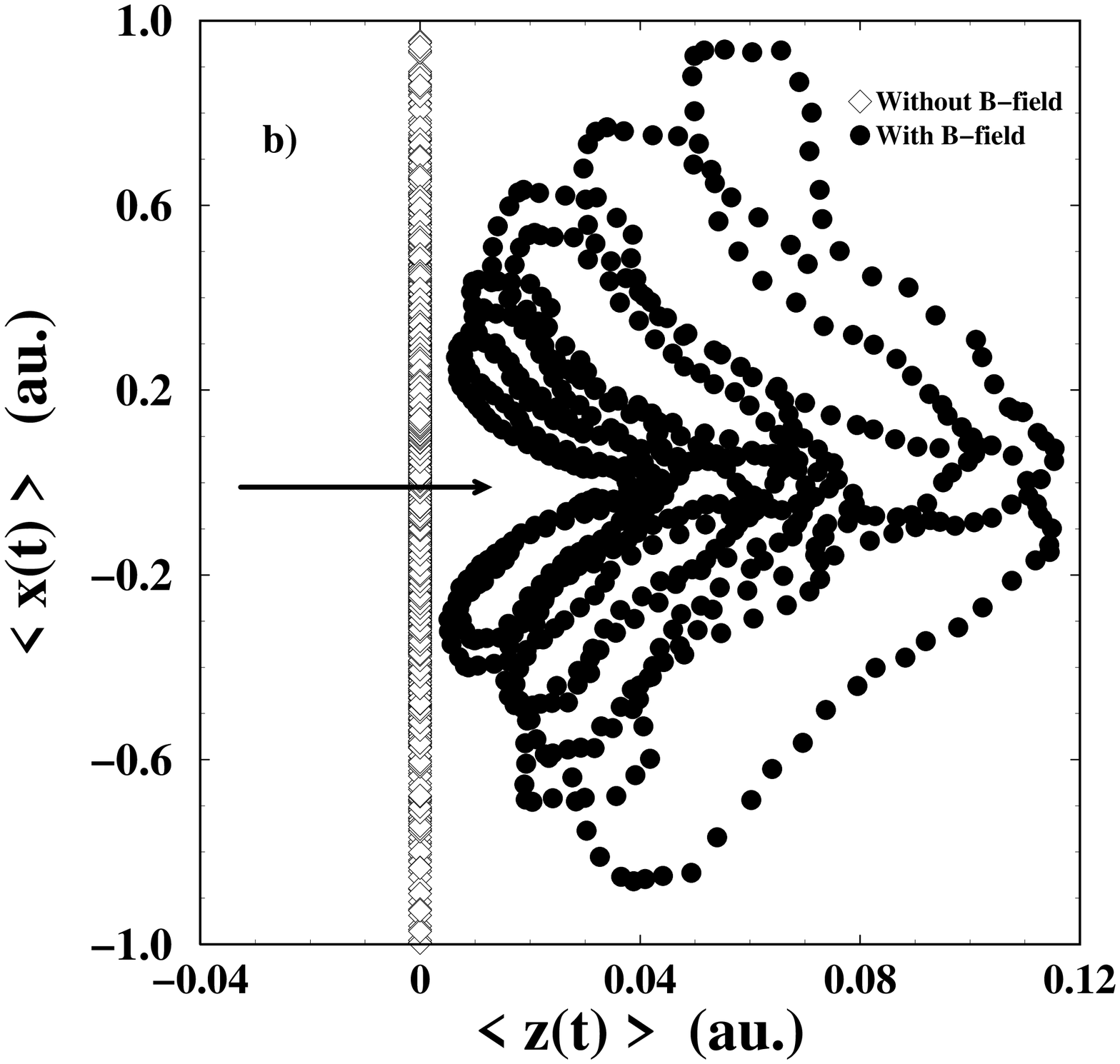}} 
\end{center} 
%\vspace{1cm} 
\end{figure} 
 
Fig. 3b: S. X. Hu  and C. H. Keitel, ``Dynamics of ...'' 

\newpage

\begin{figure} 
\begin{center} 
\unitlength1cm 
   \makebox {\epsfysize 18cm \epsffile{hufig4.ps}} 
\end{center} 
%\vspace{1cm} 
\end{figure} 
 
Fig. 4: S. X. Hu  and C. H. Keitel, ``Dynamics of ...'' 

\newpage

\begin{figure} 
\begin{center} 
\unitlength1cm 
   \makebox {\epsfysize 15cm \epsffile{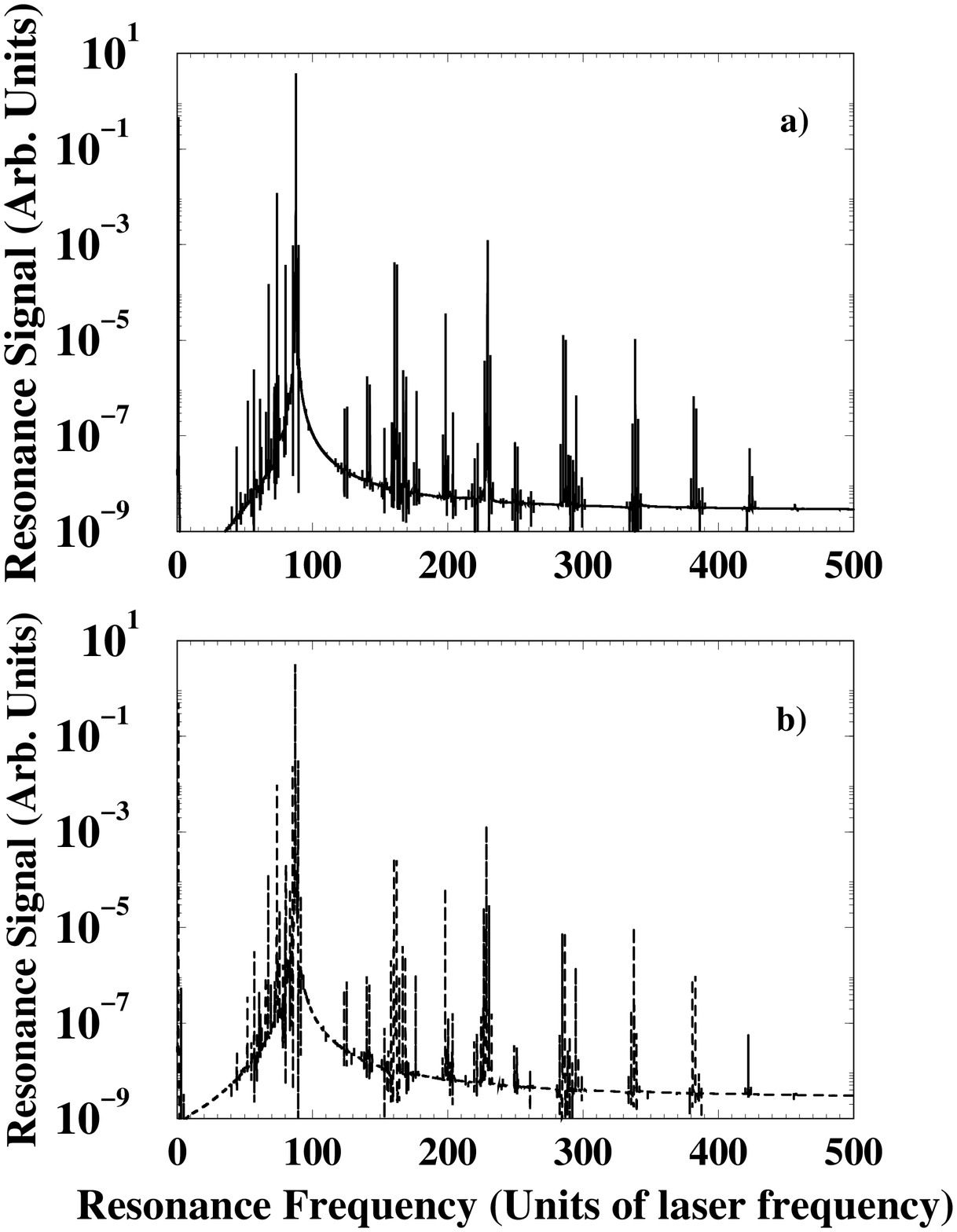}} 
\end{center} 
%\vspace{1cm} 
\end{figure} 
 
Fig. 5: S. X. Hu  and C. H. Keitel, ``Dynamics of ...'' 

\newpage

\begin{figure} 
\begin{center} 
\unitlength1cm 
   \makebox {\epsfysize 15cm \epsffile{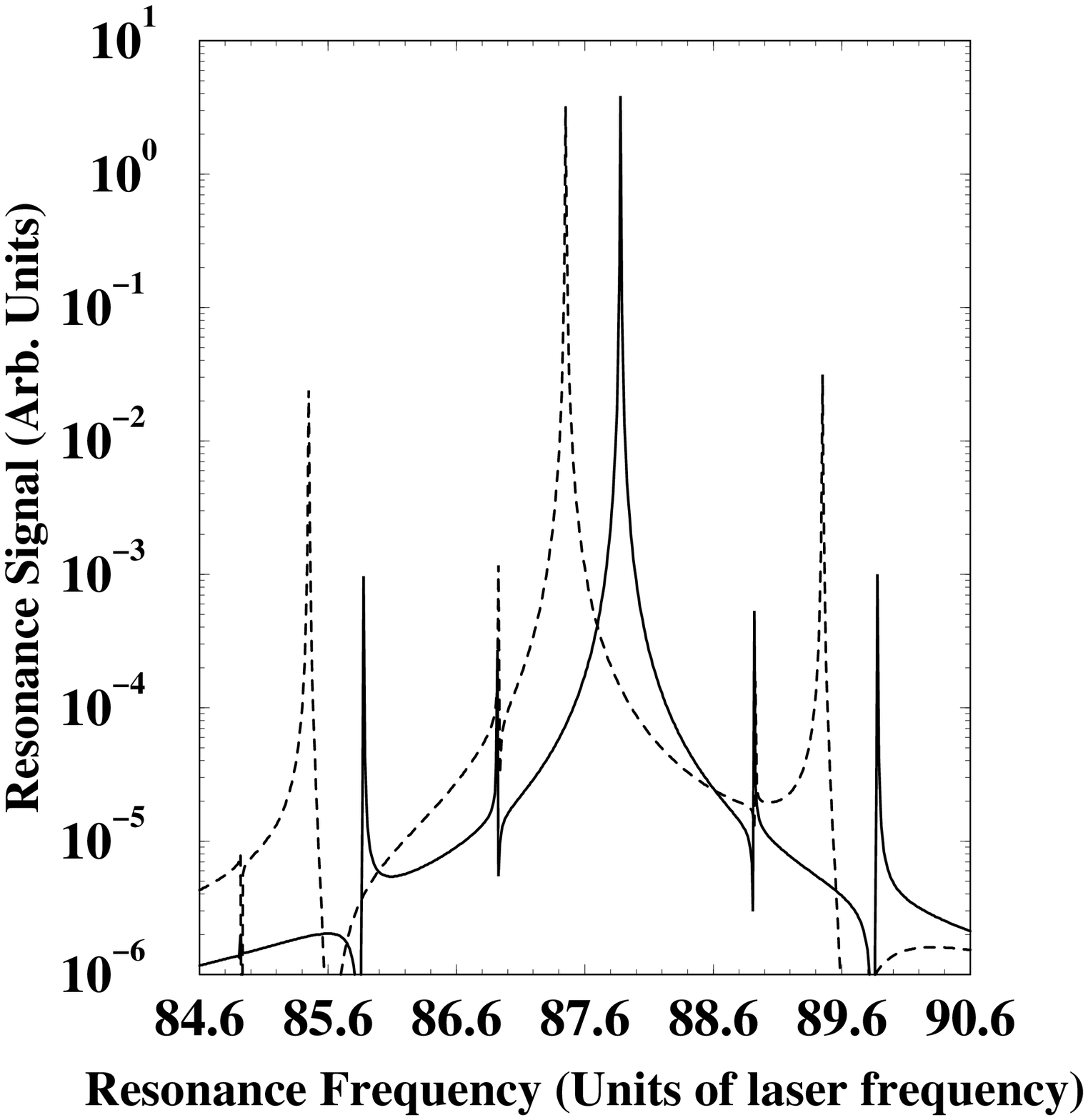}} 
\end{center} 
%\vspace{1cm} 
\end{figure} 
 
Fig. 6: S. X. Hu  and C. H. Keitel, ``Dynamics of ...'' 

\newpage

\begin{figure} 
\begin{center} 
\unitlength1cm 
   \makebox {\epsfysize 15cm \epsffile{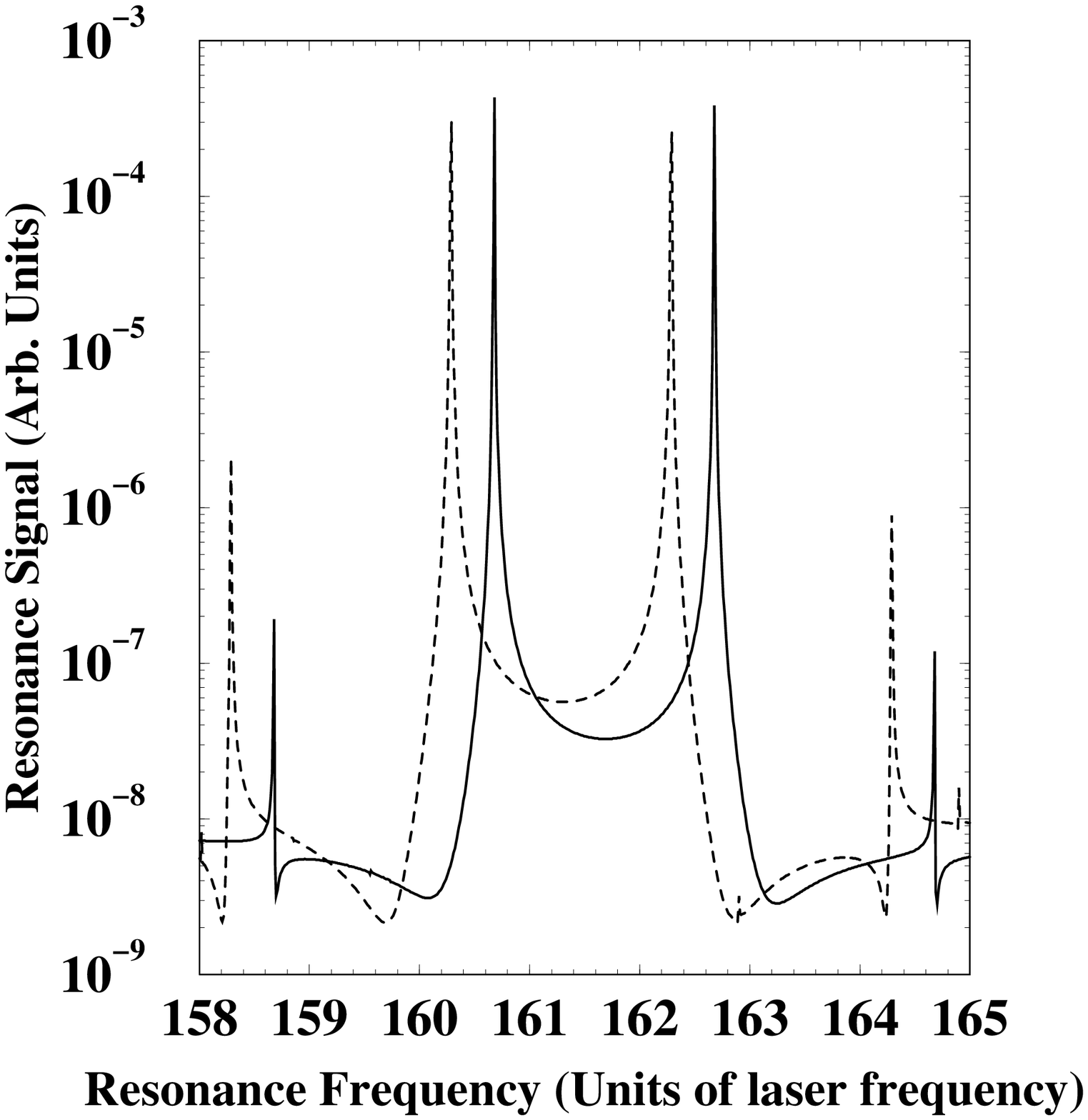}} 
\end{center} 
%\vspace{1cm} 
\end{figure} 
 
Fig. 7: S. X. Hu  and C. H. Keitel, ``Dynamics of ...'' 

\newpage

\begin{figure} 
\begin{center} 
\unitlength1cm 
   \makebox {\epsfysize 15cm \epsffile{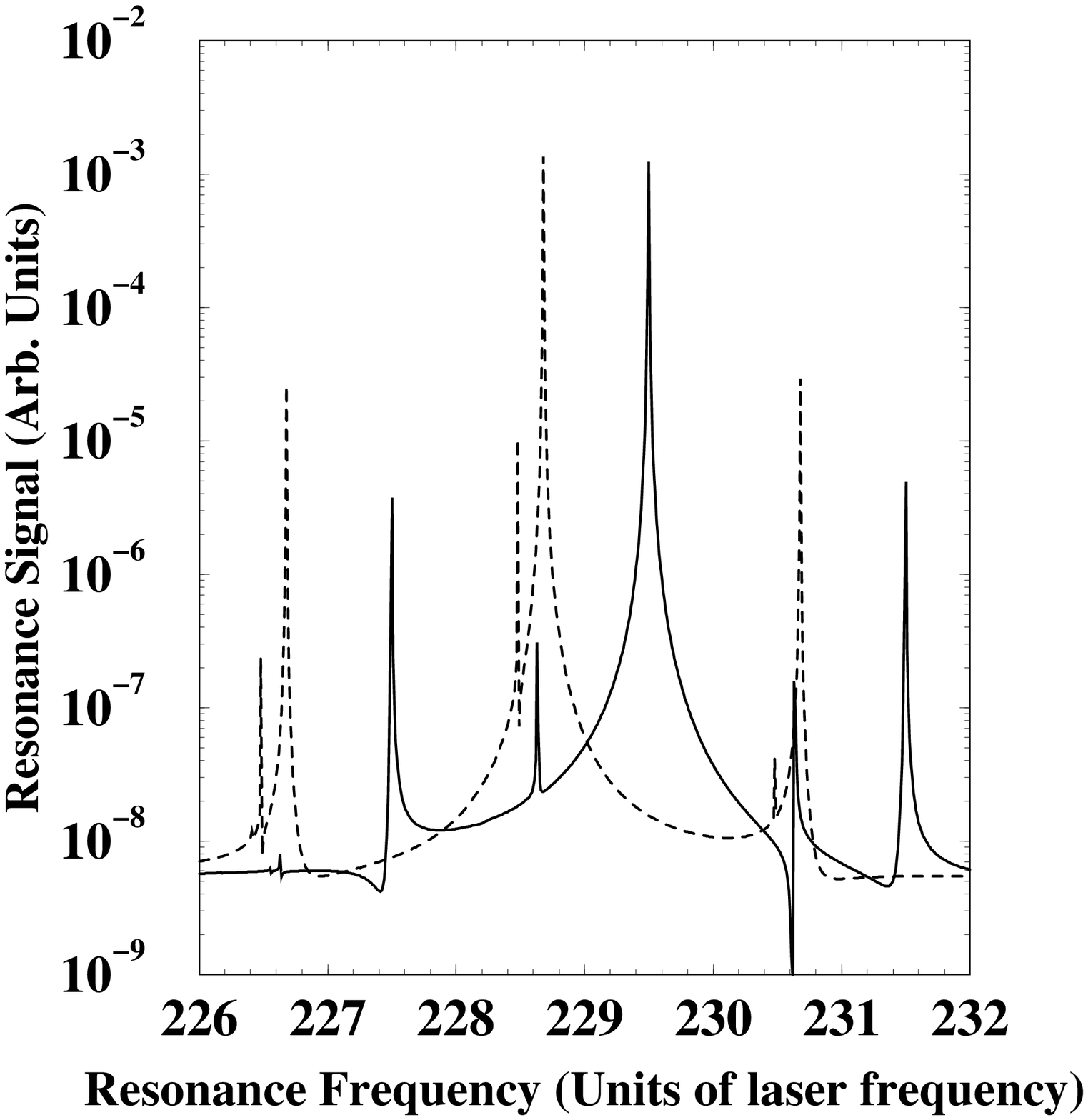}} 
\end{center} 
%\vspace{1cm} 
\end{figure} 
 
Fig. 8: S. X. Hu  and C. H. Keitel, ``Dynamics of ...'' 

\newpage

\begin{figure} 
\begin{center} 
\unitlength1cm 
   \makebox {\epsfysize 15cm \epsffile{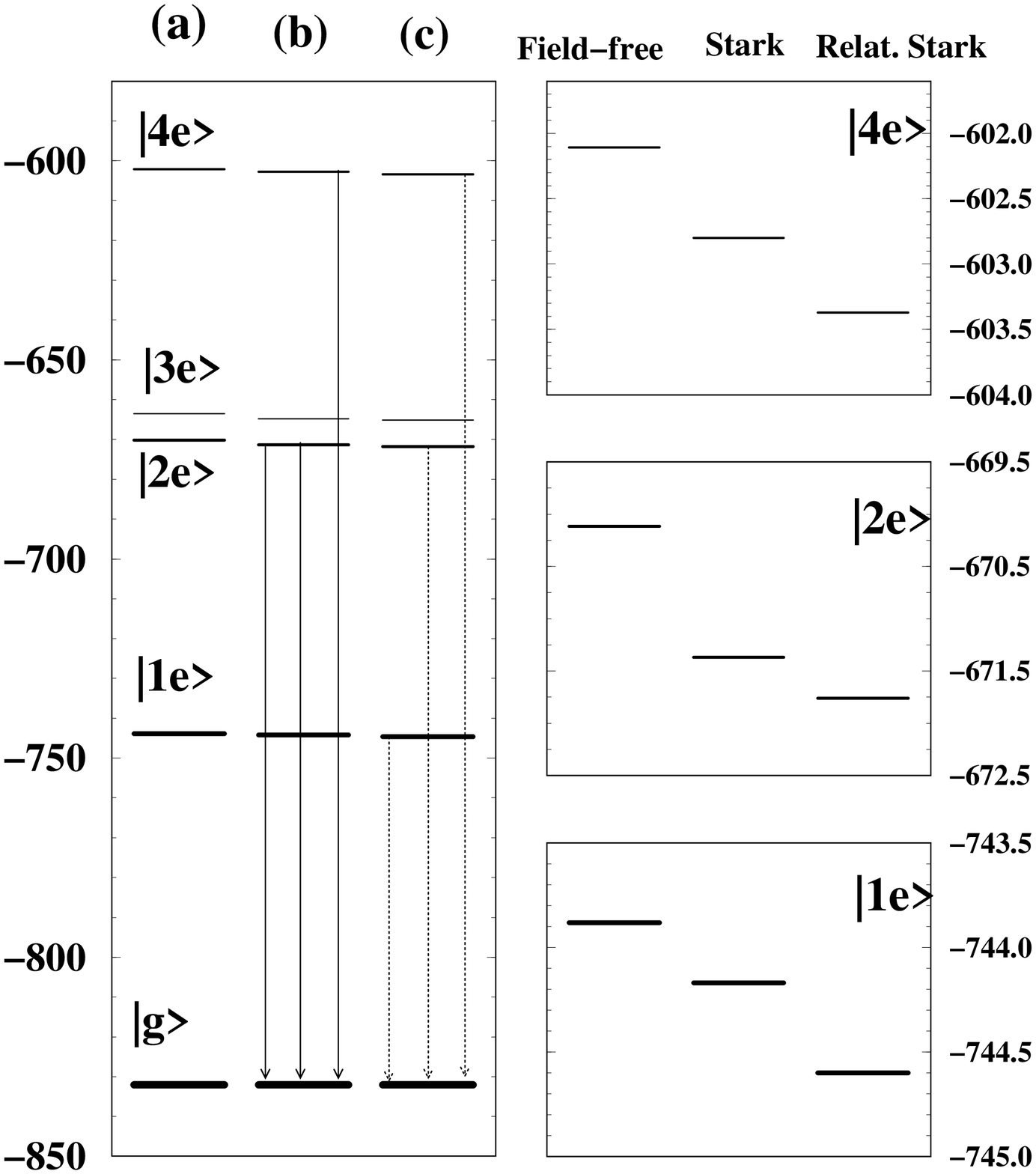}} 
\end{center} 
%\vspace{1cm} 
\end{figure} 
 
Fig. 9: S. X. Hu  and C. H. Keitel, ``Dynamics of ...'' 

\newpage

\begin{figure} 
\begin{center} 
\unitlength1cm 
   \makebox {\epsfysize 15cm \epsffile{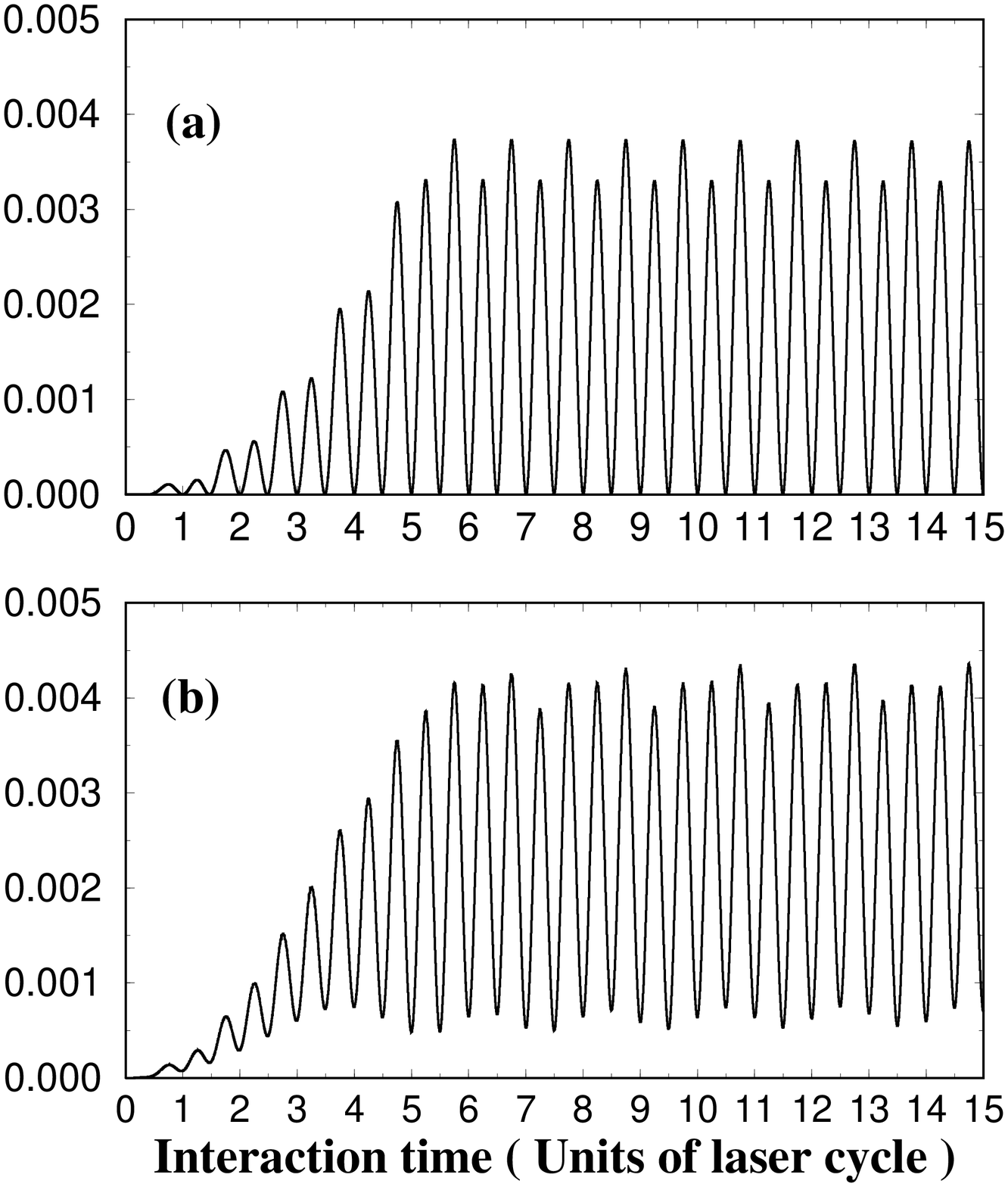}} 
\end{center} 
%\vspace{1cm} 
\end{figure} 
 
Fig. 10: S. X. Hu  and C. H. Keitel, ``Dynamics of ...'' 

\newpage

\begin{figure} 
\begin{center} 
\unitlength1cm 
   \makebox {\epsfysize 20cm \epsffile{hufig11.ps}} 
\end{center} 
%\vspace{1cm} 
\end{figure} 
 
Fig. 11: S. X. Hu  and C. H. Keitel, ``Dynamics of ...'' 

\newpage

\begin{figure} 
\begin{center} 
\unitlength1cm 
   \makebox {\epsfysize 15cm \epsffile{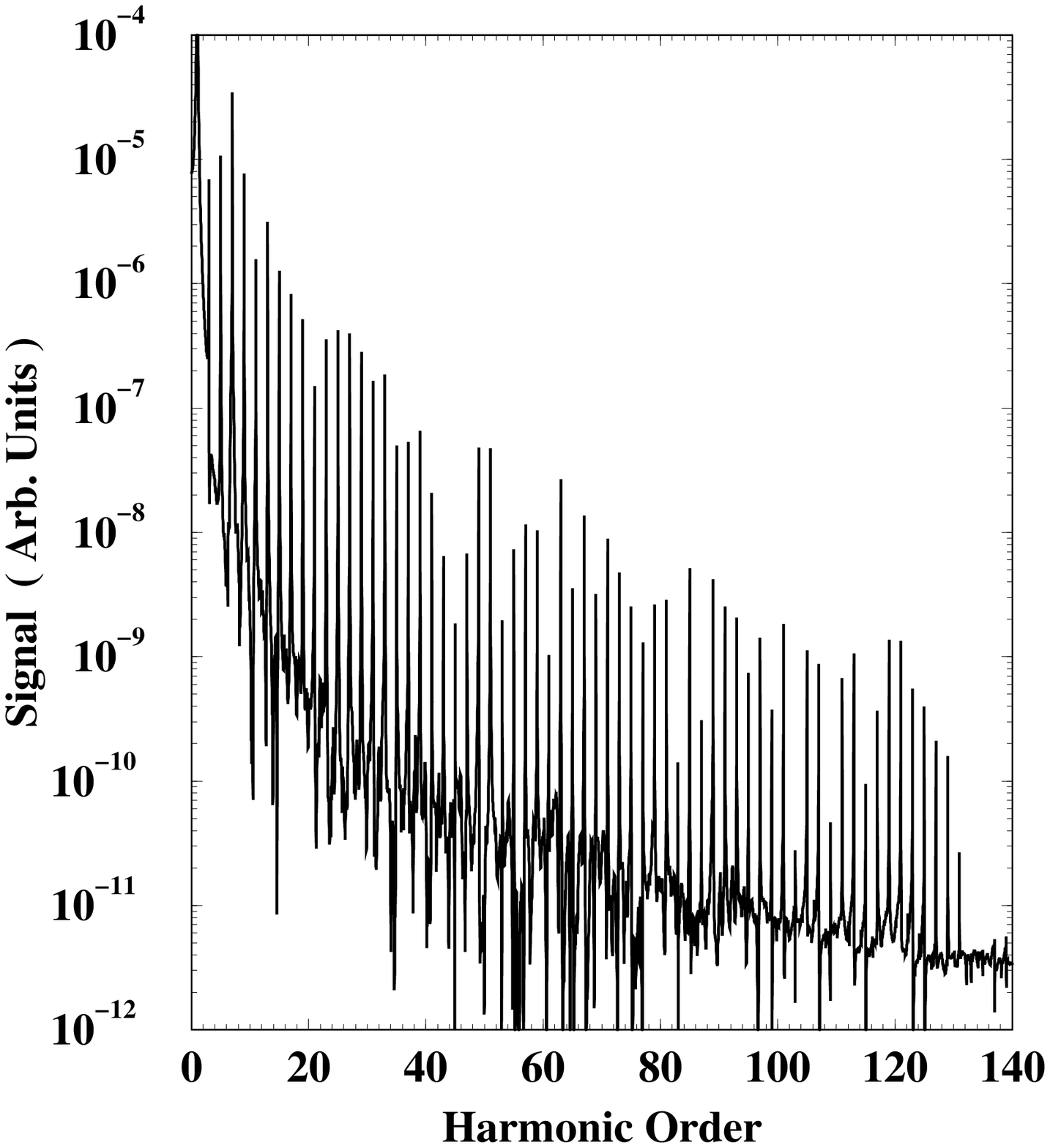}} 
\end{center} 
%\vspace{1cm} 
\end{figure} 
 
Fig. 12: S. X. Hu  and C. H. Keitel, ``Dynamics of ...'' 

\newpage

\begin{figure} 
\begin{center} 
\unitlength1cm 
   \makebox {\epsfysize 15cm \epsffile{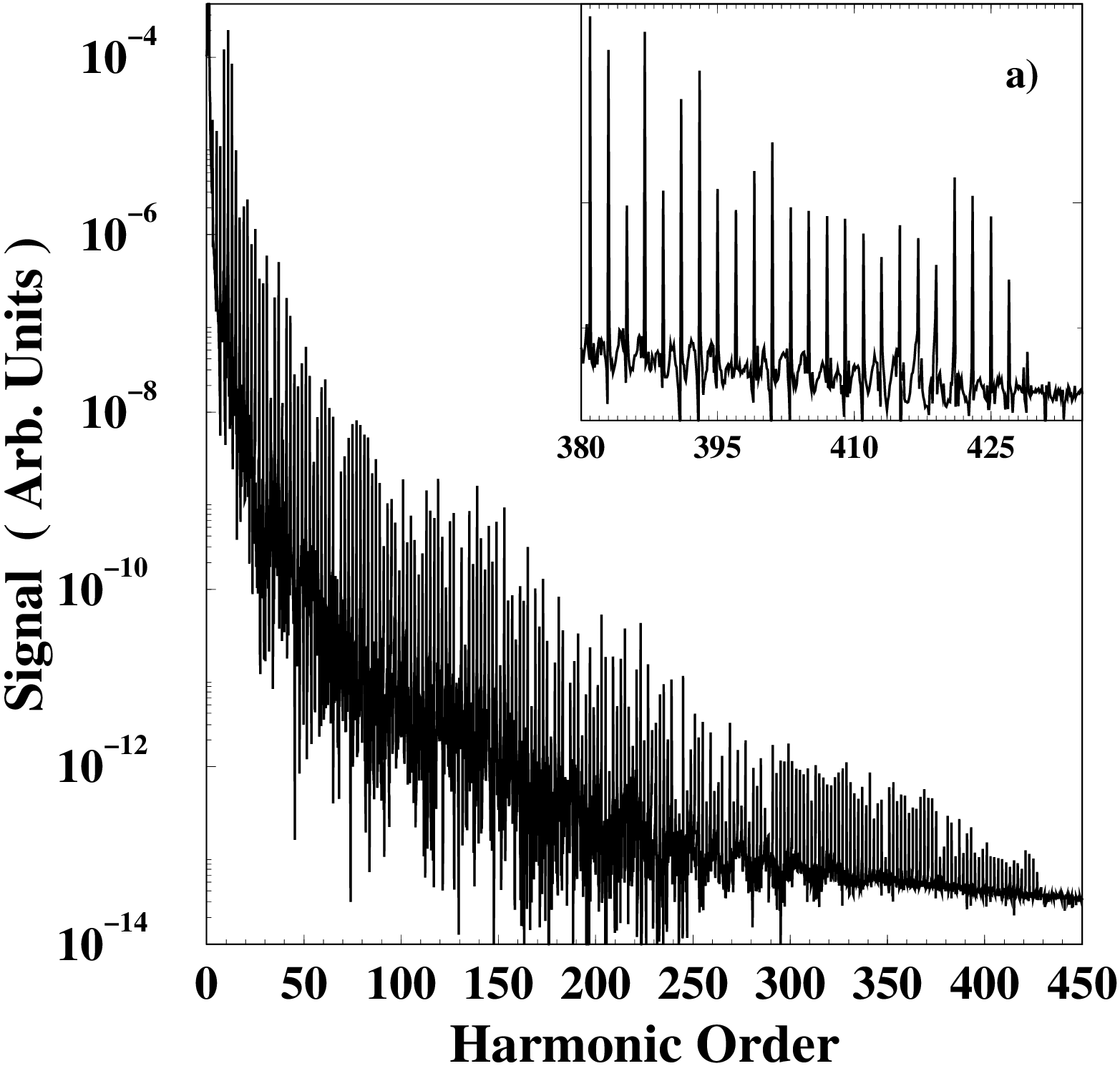}} 
\end{center} 
%\vspace{1cm} 
\end{figure} 
 
Fig. 13a: S. X. Hu  and C. H. Keitel, ``Dynamics of ...'' 

\newpage

\begin{figure} 
\begin{center} 
\unitlength1cm 
   \makebox {\epsfysize 15cm \epsffile{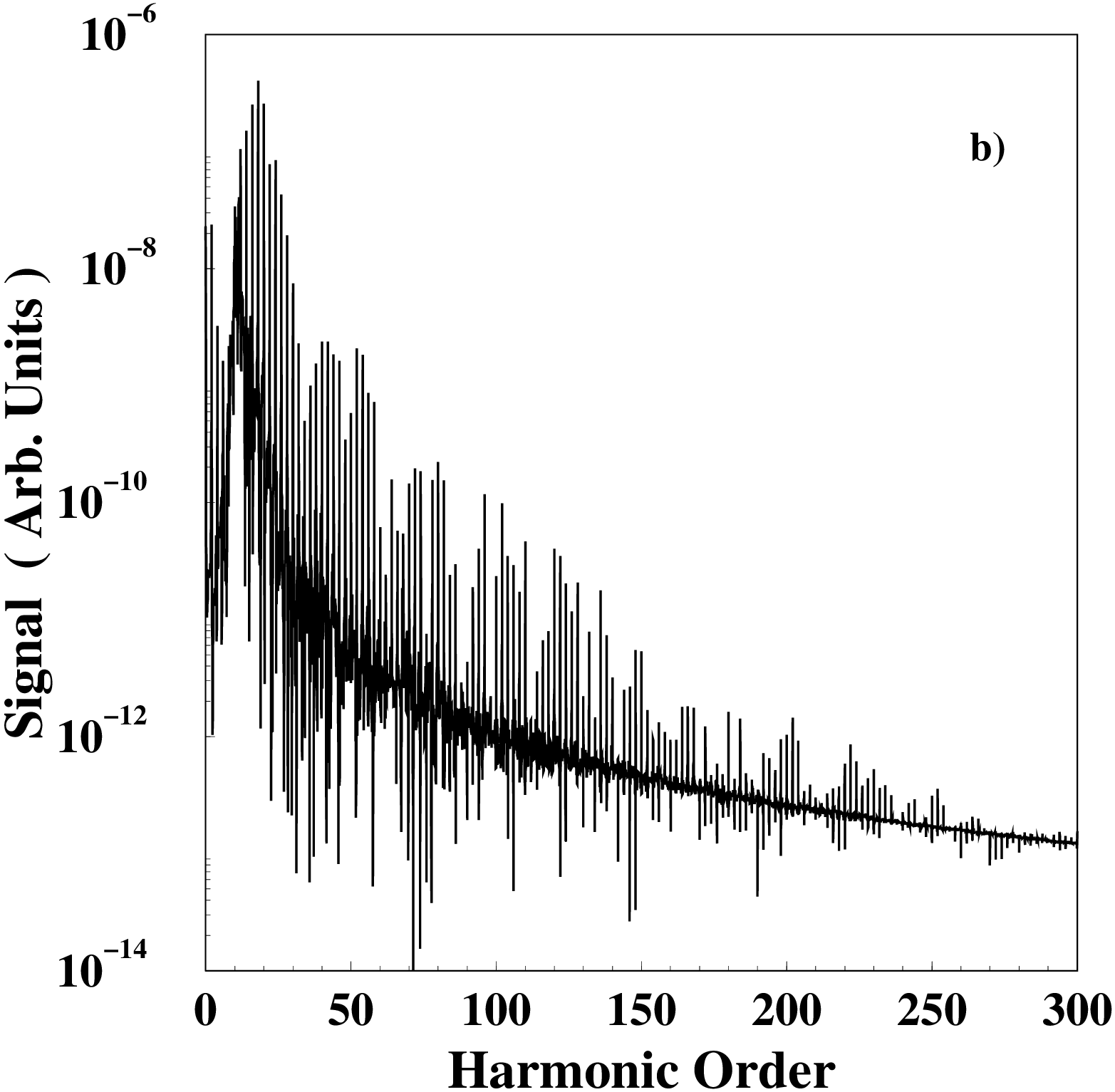}} 
\end{center} 
%\vspace{1cm} 
\end{figure} 
 
Fig. 13b: S. X. Hu  and C. H. Keitel, ``Dynamics of ...'' 

\newpage

\begin{figure} 
\begin{center} 
\unitlength1cm 
   \makebox {\epsfysize 15cm \epsffile{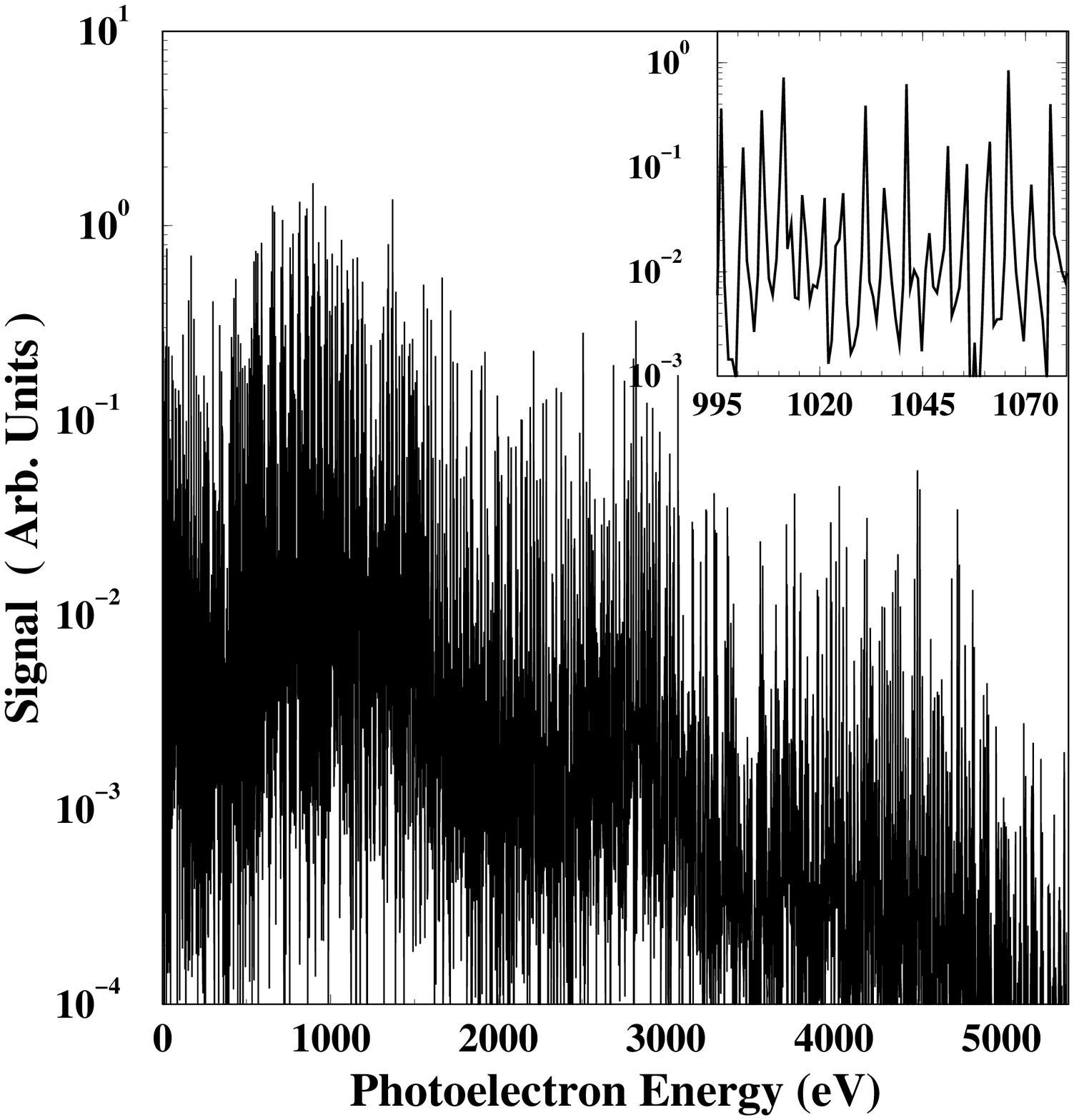}} 
\end{center} 
%\vspace{1cm} 
\end{figure} 
 
Fig. 14: S. X. Hu  and C. H. Keitel, ``Dynamics of ...'' 

\newpage

 \begin{figure} 
\begin{center} 
\unitlength1cm 
   \makebox {\epsfysize 15cm \epsffile{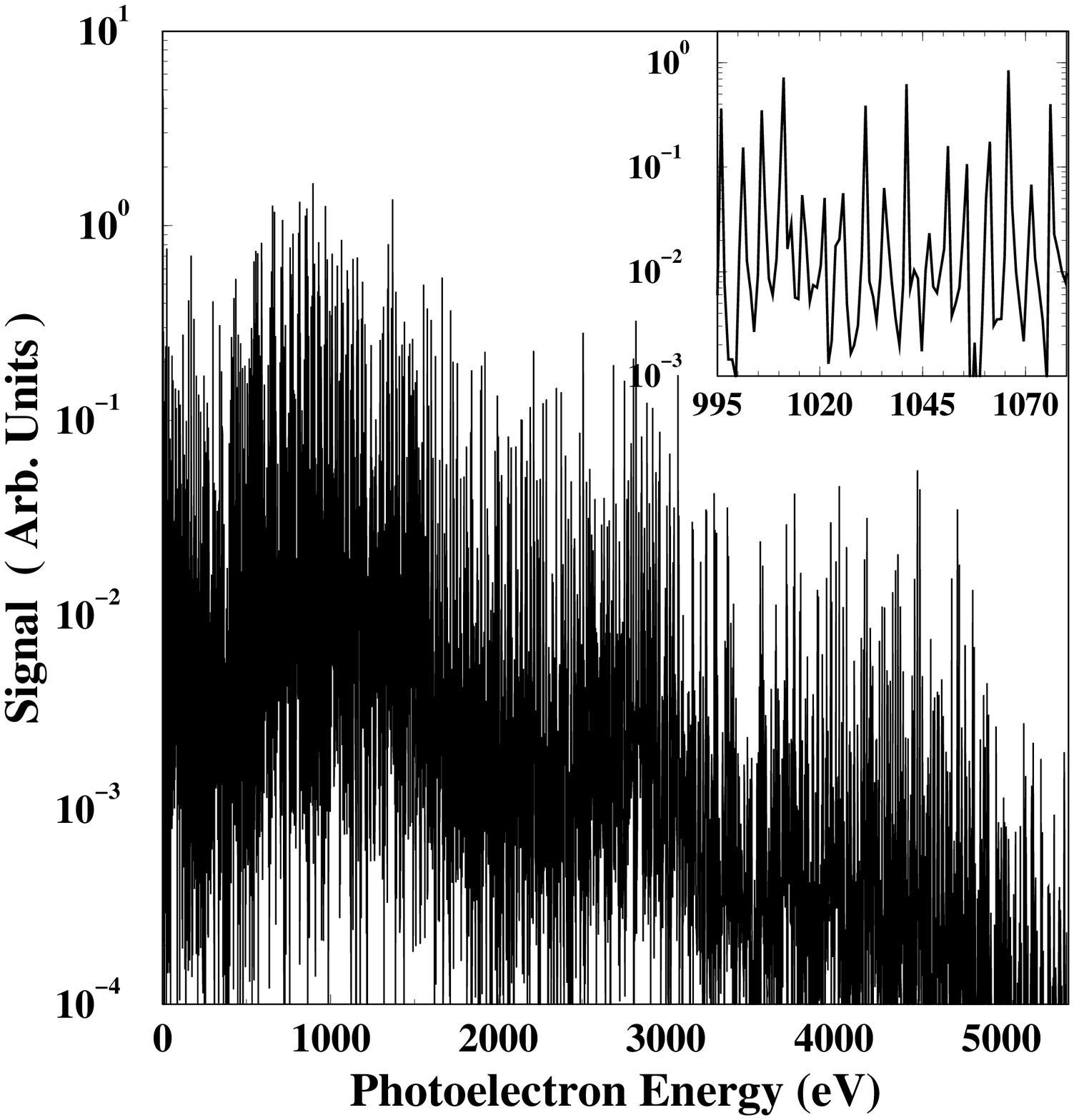}} 
\end{center} 
%\vspace{1cm} 
\end{figure} 
 
Fig. 15: S. X. Hu  and C. H. Keitel, ``Dynamics of ...'' 

\end{document}